\documentclass[authoryear]{elsarticle}
\usepackage{caption}
\usepackage{subcaption}
\usepackage[english]{babel}
\usepackage{amsmath}
\usepackage{amsfonts}
\usepackage{multirow}
\usepackage{tablefootnote}
\usepackage{amssymb}
\usepackage{hyperref}
\usepackage{algpseudocode}
\usepackage{algorithm}
\usepackage[table]{xcolor}
\usepackage{enumitem}
\hypersetup{
    colorlinks=true,
    linkcolor=blue,
    filecolor=magenta,      
    urlcolor=cyan,
}
 
\usepackage{lineno}
\modulolinenumbers[1]

\journal{Ecological Modelling}

\begin{document}
\begin{frontmatter}
\title{Statistical methods for estimating ecological breakpoints and prediction intervals}
\author[mymainaddress]{Jabed H Tomal\corref{mycorrespondingauthor}}
\cortext[mycorrespondingauthor]{Corresponding author}
\ead{jtomal@tru.ca}
\author[mysecondaryaddress,mycurrentaddress]{Jan JH Ciborowski}
\address[mymainaddress]{Department of Mathematics and Statistics, Thompson Rivers University, 805 TRU Way, Kamloops, BC, Canada V2C 0C8.}
\address[mysecondaryaddress]{Department of Biological Sciences, University of Windsor, 401 Sunset Avenue, Windsor ON, Canada, N9B 3P4.}
\address[mycurrentaddress]{Present Address: Department of Biological Sciences, University of Calgary, 
2500 University Drive NW,
Calgary, AB, Canada, T2N 1N4.}

\begin{abstract} 

The relationships among ecological variables are usually obtained by 
fitting statistical models that go through the conditional means of the dependent
variables. For example, the nonparametric loess regression model and the 
parametric piecewise linear regression model, which pass through the 
conditional mean of the response variable given the predictor, are used to 
analyze simple to complex relationships among variables. 
We  used locally estimated scatterplot smoothing 
(loess; and bootstrapped confidence intervals) to subjectively identify 
the
number and positions of potential ecological breakpoints 
in a bivariate relationship, and a
piecewise linear regression model (PLRM) to quantitatively estimate 
the location of breakpoints and the associated precision. 
We also estimated breakpoint location and precision using a
piecewise linear quantile regression model (PQRM), which is  fitted to 
the quantiles of the
conditional distribution of the response variable 
given the predictor and provides
much richer information in terms of estimating relationships and breakpoints.
We compared the precision of breakpoints estimated by PQRM relative to PLRM.
In the environmental
literature, bootstrapped loess, piecewise linear regression and single-breakpoint 
PQRM have all been proposed to generate
prediction bands for an ecological response against human induced 
disturbances
in nature. 
We compared the precision of  the methods using two examples 
from the ecological
literature suspected to exhibit multiple breakpoints: 
relating a Fish Index of Biotic Integrity 
(an index of wetlands'  fish community `health') to the amount
of human activity in wetlands' adjacent watersheds; and relating 
the biomass 
of cyanobacteria to the total phosphorus concentration in Canadian lakes. 
Statistically significant breakpoints were detected for both datasets, 
demarcating the boundaries of \emph{three} line segments with markedly different slopes.
PQRM generated confidence limits of the breakpoints that were consistently 
narrower than limits estimated by the PLRM. 
Similarly, the prediction bands of the relationships estimated by PQRM were 
only $72.15\%$ to $74.60\%$ as wide as the bands estimated using PLRM.    
We recommend the  piecewise linear quantile regression as an effective 
means of characterizing bivariate environmental relationships where the
scatter of points represents natural environmental variation rather 
than measurement error.

\end{abstract}

\begin{keyword}
Ecological breakpoints \sep
bootstrap \sep
loess \sep
piecewise linear regression \sep
quantile regression \sep
chlorophyll \sep
phosphorus \sep
index of biological integrity \sep
environmental stress
\end{keyword}

\end{frontmatter}


\section{Introduction} 

Maintaining the health of aquatic habitat (biotic integrity) is one of the primary objectives
set forth by the US Clean Water Act of $1972$ (PL $92$ - $500$).
``Biotic integrity'' is defined
as the ``ability of a habitat to support and maintain a balanced,
integrated, adaptive community of organisms having a
composition, diversity and function
comparable
to that of a natural habitat \citep{Frey:1977}.'' 
The global pervasiveness of human activity means that few
or no areas can be considered as completely ``natural'' habitat.
Instead, comparisons are commonly made to locations that are
in the reference condition, referring to an area subject to a minimum
level of anthropogenic stress \citep{Karr:1998,Host:2005,Stoddard:2006}.
Questions of identifying the degree of disturbance (`numerical criterion')
at which biological changes occur have long been an important consideration
\citep{Karr:1998,KILGOUR:1998,Qian:2003,Qian:2015}.
Such values can serve as guidelines used for the protection 
of environmental integrity of sites (by preventing or mitigating activities
that would result in criteria being exceeded), and as theoretical restoration targets \citep{Johnson:2013}. 
Statistical approaches and methods of identifying 
such environmental disturbance thresholds have been a topic of considerable research.
\cite{Bunea:1999} concluded that regression trees could be an 
effective means of dividing a sequence of biological values 
ordinated across a stress gradient into multiple classes representing tiers of ecological condition.
\cite{Qian:2003} proposed a nonparametric approach and a Bayesian approach
to identifying the location of change-points in the environmental stress-ecological condition relationship.
\cite{Brenden:2008} evaluated the utility of several model-based approaches to identify single 
disturbance thresholds, including nonparametric deviance 
reduction (NDR [=regression trees]), piecewise regression (PR), 
Bayesian change-point (BCP), quantile piecewise constant (QPC), and 
quantile piecewise linear (QPL) approaches using simulated data. 
They concluded that QPL approaches most consistently identified the environmental 
change-point at which variation in a univariate dependent variable was altered. 
They considered stair-step, condition mean, and wedge response patterns. 
\cite{Dodds:2010} reviewed additional methods commonly employed to 
detect single-inflection point nonlinearities in stress-response relationships. 
These models are appropriate for identifying single change-points at which 
the measure of central tendency of a population of response measurements changes.   

However, such models are limited in two perspectives. 
Firstly, they assume that the bivariate relationship is 
based upon the association of two simple variables 
drawn from a single population of values. Yet, in 
bioassessment the dependent variable is often a 
composite measure (such as an Index of 
Biological Integrity \citep[IBI;][]{Karr:1998}
estimated from samples of many different systems, 
each with differing component properties. 
This renders identification of the precise location of 
change-points difficult \citep{King:2011}. 
Secondly, when the independent variable is the only one 
of many potentially limiting factors, which interact to influence the dependent variable,
the observed relationships can be weak or nonsignificant \citep{Cade:1999,Cade:2003}.
Such interactions produce the characteristic wedge-shaped pattern often 
observed when plotting simple bivariate relationships whereby the dependent 
variable is limited by the effects other unmeasured independent variables \citep{Cade:1999,Cade:2005}.
Quantile regression has the potential to accommodate these limitations.
\cite{Scharf:1998} found that quantile regression was the most robust
of several approaches to identifying functions to characterize the upper and lower
boundaries of response variables. \cite{Horning:2012} used an application
of quantile regression (Kandall-Theil robust line estimation \citep{Granato:2006}) to
derive a nonlinear constraint envelope bounding the relationship between duration
of marine mammal dives and time between dives.
In this paper, we contrast three regression model approaches - two based on central tendency and one
based on overall distribution of the response variable. The questions we address are:
$a$: is piecewise quantile regression better able to detect nonlinearities or the presence of a threshold
than central tendency models typically employed?; and $b$: are threshold estimates generated by piecewise quantile
regression more precise than other models? We apply the procedures to two data sets relating ecological condition
of North American aquatic systems and inferred effects of stress variables on biota.

The first application relates to assessing/estimating threshold effects of agricultural
activity in catchments (watersheds) draining into the Laurentian Great Lakes and an index
of community composition of fishes in bordering coastal wetlands. Run-off associated with agriculture
is a major source of human induced disturbance affecting natural habitat loss for fishes in lakes and 
rivers \citep{Brazner:1997,Crosbie:1999}.
In this case, the measure of ecological condition is a wetland fish index of 
biotic integrity \citep[Fish IBI;][]{Uzarski:2005,Bhagat:2007},
a metric representing the health of an ecoregion or watershed. \cite{Danz:2005} 
derived a composite agricultural
stress gradient to characterize risk of degradation of natural habitat using 
Geographic Information System (GIS) based data.
\cite{Uzarski:2005} developed a fish multimetric index of biotic integrity through assessing fish community
composition in wetland areas where bulrush (\textit{Schoenoplectus, spp}) was the 
dominant vegetation across the US and Canada Great Lakes Coastline.

The second application is concerned with identifying threshold concentrations of total
phosphorus - that increase the risk of harmful algal blooms dominated by potentially toxigenic Cyanobacteria in lakes
\citep{Watson:1992,Downing:2001}. Total phosphorus (TP) is a limiting nutrient whose loads to lakes and rivers
reflect contributions of sewage from urban centres, agricultural runoff and other manifestations of human
activity \citep{Reynolds:1975,Qian:2003}. Cyanobacterial blooms can contribute to hypolimnetic oxygen depletion,
impacts on recreation, ecosystem integrity, human and animal health, and their toxic byproducts can overwhelm the capacity
of municipal water treatment plants to provide safe drinking water \citep{Downing:2001,Beaulieu:2014}.

Opinion on the shape of the relationship between total phosphorus and cyanobacteria biomass is varied. Total
phosphorus is arguably the best single predictor of cyanobacterial biomass (sometimes measured
as Chlorophyll a), and empirically-derived linear models are widely used in 
lake management \citep[e.g.,][]{Dillon:1974,Dillon:1975,Stow:2013,Beaulieu:2014}.
However, sigmoidal relationships between total phosphorus and the biomass or relative 
abundance of cyanobacteria are well 
documented \citep{Watson:1992,Chow-Fraser:1994,Watson:1997,Downing:2001,Filstrup:2014}. 
The management implications associated with applying these
two different interpretations of the relationship are significant. 
Use of a linear model to guide management implies that any alteration
in total phosphorus concentration in a receiving water body can be expected to result 
in a cyanobacterial response. In contrast, adherence to a sigmoidal
model implies that there are points of inflection beyond which the 
two variables may behave independently, possibly obviating the need
for the control of total phosphorus below a particular concentration.

Traditionally, points of inflection of potentially nonlinear models 
have been identified using
loess, logistic or median regression analyses, which track the 
central tendency of the relationship to reveal overall behavior. 
However, if goals are to apply a precautionary
principle to nutrient management, inflection points should be 
based on the boundaries (prediction limits) of the data rather 
than their maximum likelihood of occurrence. We
investigated the effectiveness of using piecewise quantile 
regression as a means of identifying these points.

\cite{Beaulieu:2014} summarized
data from
$149$ Canadian lakes
collected by Ministries of the Environment of Alberta, British Columbia,
and Ontario, Canada. Using linear and nonlinear regression and mixed-effects models,
they found that phosphorus and nitrogen were the best predictors of 
cyanobacterial biomass, but concluded that linear models better explained the data pattern
than nonlinear approaches. Yet, scatterplots of their data appear to indicate
nonlinearities at the upper and lower quantiles of the functions.

An ecological threshold is defined as a point of abrupt change of the response
variable of an ecological process (such as an index of ecological health)
against a measure of habitat, such as the human induced disturbance affecting
natural habitat \citep{Fahrig:2001,Francesco:2009}. Ecological thresholds
are useful in diagnosing the condition of a natural habitat, and may provide guidelines
for setting conservation targets \citep{Johnson:2013}. Hence it is essential to identify
a point at which human induced disturbance alters natural habitat to the extent
that a significant change in the response variable occurs.

Nonparametric regression is a frequently used and effective means of studying
simple to complex relationships between a dependent and an independent variable.
\cite{Trexler:1993} discussed the application of locally estimated
scatterplot
smoothing (loess) in ecology. Building on the recommendations of \cite{Toms:2003},
we propose using loess to subjectively identify the number and positions of ecological breakpoints
along a stressor gradient.

Piecewise linear regression is a popular approach to estimating the location of ecological thresholds.
\cite{Shea:2002} developed piecewise linear regression methodology for identifying discontinuities (thresholds)
in measurements of ecological variables. 
\cite{Toms:2003} subsequently demonstrated its value in modeling ecological thresholds, and compared a sharp-transition model
with three models incorporating smooth transitions: the hyperbolic-tangent, bent-hyperbola,
and bent cable models. \cite{Ficetola:2009} compared a wide variety
of statistical models, and suggested that that piecewise linear regression
model is an especially effective means of identifying ecological thresholds. In
all of the mentioned applications, the authors, used the piecewise linear
regression model to identify single breakpoints. In this paper,
we extended the model to incorporate two breakpoints and document its application
to the two studies described above.

We complement the use of these two well-recognized approaches with a novel
application of quantile regression. Quantile regression is a method of estimating
relationships between variables in ecological process defined through different quantiles
of the conditional distribution of the response variable. As a result,
quantile regression models provide a more complete view of possible causal
relationships between variables.
\cite{Cade:2003} gave a general overview of ecological applications of quantile regression.
They discussed linear and non-linear regression models with both homogeneous
and heterogeneous error variances. \cite{McClain:2001} studied the relationship
between dissolved oxygen concentration and maximum size in deep-sea turried gastropods using a linear quantile regression model. 
\cite{Austin:2007}
reviewed the ecological applications of a number of statistical methods
including the incorporation of linear and nonlinear quantile regressions into
species response models used in conservation. \cite{Bissinger:2008}
predicted marine phytoplankton maximum growth rates from temperature
using a non-linear quantile regression model. \cite{Planque:2008} used a
linear quantile regression model to study fish recruitment-environment
relationships in marine ecology. \cite{Cade:2005} used linear and non-linear
quantile regression models to reveal hidden bias and uncertainty in habitat models
in ecology. \cite{Brenden:2008} evaluated piecewise quantile regression
as means of detecting a single disturbance threshold. Similarly, in this
study, we used a piecewise linear quantile regression model to estimate
ecological thresholds. The estimates of a threshold for different quantiles provide
a precise means of identifying confidence intervals for the breakpoints. 
The
confidence interval of a threshold provides an idea of the accuracy in estimation
of the thresholds.

Construction of prediction intervals of the response variable given the predictor
can also be informative as illustrated by
\cite{Karanth:2004}, who generated
prediction
intervals for tiger densities as a function of their prey densities. 
In this paper,
we
assess the precision of prediction intervals determined by three methods
for each of two ecological stress-response relationships - a wetland fish community
index of biotic integrity relative to the extent of agricultural activity in contributing
drainage basins; and cyanobacteria biomass in lakes relative to total phosphorus concentration.
We construct three prediction intervals using three models:
 nonparametric regression loess, piecewise linear
regression, and piecewise linear quantile regression.

We first describe the statistical
methods used  for each of the three approaches: loess and bootstrap prediction band,
piecewise linear regression model, and piecewise linear quantile regression model. 
Subsequently, we apply the methods to
$(a)$  wetland fish
index of biotic integrity versus agricultural stress, and the  $(b)$ cyanobacteria biomass
versus total phosphorus data sets. We then assess the relative performance of the approaches 
and their implications for interpreting the ecological data sets.

\section{Methods} \label{sect:methods}

\subsection{Summary of models}

\subsubsection{Loess and bootstrap prediction band}
Let $y$ and $x$ be the response and predictor variables, respectively.
A nonparametric smoothed regression model 
\citep{Cleveland:1979} is defined
as
\begin{equation} \label{eqn:loess}
y = m(x; h) + \epsilon,
\end{equation}
where $m(x; h)$ is the smoothed function of interest with smoothing parameter $h$
and $\epsilon$ is an independent error term with mean $0$ and constant scale $\sigma^2$.
The smoothed function $m(x; h)$ is obtained by fitting a polynomial using weighted
least squares with varying large and small weights for the nearby and distant observations,
respectively, in a neighborhood of $x$.
This model is known as locally estimated scatterplot smoothing, loess.
The loess model is robust
against a few outliers in the data
and prevents extreme observations from exerting a large influence in the fitting
procedure.

The nonparametric smoothed regression model captures both linear and
nonlinear relationships among variables. When the linear models fit poorly
due to intrinsic non-linearity in the data, the nonparametric smoothed
regression model appears to be effective in identifying 
relationships among variables. 
Our purpose  of  fitting a nonparametric smoothed regression
model, is to identify the number and
position(s) of the possible breakpoints(s) in ecological relationships.

When a nonparametric regression model is used to visualize
the relationships between variables, generation of a  prediction
band provides a measure of the variability of the response variable $y$
given 
a particular value of
the predictor variable $x$.
Bootstrap samples \citep{Efron:1994} can be used to generate the nonparametric prediction band for the
ecological response.
Similar to methods presented in
\cite{Hardle:1988} and \cite{Davison:1997}, we propose
the following algorithm to construct prediction bands
for the ecological response given the predictor.

\begin{algorithm}
\small
\caption{Bootstrap resampling to construct a nonparametric prediction band for the
ecological response given the predictor.}
\begin{enumerate}[label=(\alph*)]
\item Fit loess $\hat{m}(x; h)$ of the model $m(x; h)$,
and make prediction $\hat{y}_i = \hat{m}(x_i; h)$ for $i = 1, 2, \cdots, n$.

\item Calculate $i$th residual $\hat{\epsilon}_i = y_i - \hat{y}_i$, and normalize the residuals as following $\tilde{\epsilon}_i = \hat{\epsilon}_i - \frac{1}{n} \sum_{j = 1}^n \hat{\epsilon}_j$  for $i = 1, 2, \cdots, n$.

\item For $b \; \text{in} \; 1 \; \text{to} \; B$:

\begin{enumerate}[label=(\roman*)]

\item Generate bootstrap residuals $\{\epsilon_i^*\}$ by sampling with replacement from
$\{\tilde{\epsilon}_i\}$, and calculate bootstrap observations $y_i^* = \hat{m}(x; h) + \epsilon_i^*$.

\item Fit loess $\hat{m}^*(x; h)$ by smoothing bootstrapped observations $(x_i, y_i^*)$,
and calculate bootstrapped residuals $e_i^* = y_i^* - \hat{m}^*(x_i; h)$ for $i = 1, 2, \cdots, n$.

\item Normalize the bootstrapped residuals $\tilde{e}_i^* = e_i^* - \frac{1}{n} \sum_{j = 1}^n e_j^*$ for $i = 1, 2, \cdots, n$.

\item Sample residuals $\{e_1^{**}, e_2^{**}, \cdots, e_n^{**}\}$ with replacement from the normalized bootstrapped residuals $\{\tilde{e}_1^{*}, \tilde{e}_2^{*}, \cdots, \tilde{e}_n^{*}\}$, and calculate predicted residuals
$e_i^{*p} = \hat{m}(x_i; h) - \hat{m}^*(x_i; h) + e_i^{**}$ for $i = 1, 2, \cdots, n$.

\end{enumerate}

\item End For.

\item Set the confidence coefficient $\gamma \in (0, 1)$ for the prediction band, calculate empirical quantiles $e_i^{*p}((1-\gamma)/2)$ and $e_i^{*p}((1+\gamma)/2)$ of the predicted residuals across bootstrap resamples, and construct lower and upper limits of the prediction band $\left[\hat{y}_i + e_i^{*p}((1-\gamma)/2),\ \hat{y}_i + e_i^{*p}((1+\gamma)/2)\right]$.
\end{enumerate}
\label{alg:boot:band}
\end{algorithm}

The advantage of this algorithm is that it can be applied to any parametric
or norparametric method to construct confidence bands for the response by replacing $\hat{m}(x; h)$ by appropriate model.

We used the $R$ \citep{Rcite} statement \texttt{loess} to fit the model (equation \ref{eqn:loess})
and to construct a confidence band using algorithm \ref{alg:boot:band}.
The goals are threefold: 
\begin{enumerate}
	\item[(i)] Displaying the shapes of relationships between the biological response variable and the environmental condition,
	\item[(ii)] Identifying possible locations of the breakpoints (if there are any), and 
	\item[(iii)] Determining the  prediction band for the ecological response variable given environmental condition.
\end{enumerate}

\subsubsection{Piecewise linear regression model} \label{sec:PLRM}

The simple linear regression model defines the relationship between a response
and a predictor variable. It goes through the conditional mean of the response
given the predictor and is linear in parameters and variables. A piecewise
linear regression model with one breakpoint goes through the conditional
mean of the response and connects two linear segments at the breakpoint. Similarly, a
piecewise linear regression model with two breakpoints connects three linear segments,
wherein the first and second segments are connected at the first breakpoint and the second and third linear segments are connected at the second breakpoint.

Where stress-response relationships approximate a logistic curve characterized by two inflection points, a model that identifies two breakpoints is appropriate.
We define a piecewise linear regression model
\[
\ y_i = m(x_i; \boldsymbol{\theta}) + \epsilon_i
\]
to
estimate two breakpoints incorporating three linear segments \citep{Seber:2003} as following:

\begin{equation}
m(x_i; \boldsymbol{\theta}) = \left\{\begin{array}{lcr} \beta_0 + \beta_1 x_i & \text{for} & x_i \leq \alpha_1\\
\beta_0 + \beta_1 x_i + \beta_2 (x_i - \alpha_1) & \text{for} & \alpha_1 < x_i \leq \alpha_2\\
\beta_0 + \beta_1 x_i + \beta_2 (x_i - \alpha_1) + \beta_3 (x_i - \alpha_2) & \text{for} & x_i > \alpha_2\\
\end{array} \right.
\label{eqn:PLRM}
\end{equation}
where $y_i$ and $x_i$ are the values for the $i$th response and predictor variables,
respectively,
and $\alpha_1$ and $\alpha_2$ are the two breakpoints. 
Here, the vector of parameters $\boldsymbol{\theta} = \left(\boldsymbol{\beta}^T, \boldsymbol{\alpha}^T\right)^T$
contains the vector of regression coefficients $\boldsymbol{\beta} = \left(\beta_0, \beta_1, \beta_2, \beta_3\right)^T$
and vector of break points
$\boldsymbol{\alpha} = \left(\alpha_1, \alpha_2\right)^T$.
We assume that the errors
$\epsilon_i$ are independent and identically distributed
normal random variable with zero mean, constant variance, and
finite absolute moment for some order greater than $2$. The slopes for the
first, second,
and third segments of this model are $\beta_1$, $\beta_1 + \beta_2$, and
$\beta_1 + \beta_2 + \beta_3$, respectively.
The above parametrization of the model forces continuity and abrupt
transitions at the breakpoints.
The parameters are estimated using a non-linear least squares method.
The initial values to run the non-linear least squares method are obtained
from fitting non-parametric regression loess.

We fitted the piecewise linear regression model (equation \ref{eqn:PLRM}) to the ecological response variables against their environmental predictor using the R package  \texttt{segmented}
\citep{Segmented:2015}.
The piecewise linear regression model estimates the relationship among variables and
identified the locations of the breakpoints and their confidence intervals, allowing
us to make inferences about the statistical significance of the relationships.

\subsubsection{Piecewise linear quantile regression model}

Classical regression methods focus on estimating the parameter vector $\boldsymbol{\theta}$
of a regression model $m(\mathbf{x}; \boldsymbol{\theta})$ defined at the
conditional mean of the response variable $y$ as a function of the vector of predictor
variables $\mathbf{x}$. In mathematical notation we write
$E(y|\mathbf{x}, \boldsymbol{\theta}) = m(\mathbf{x}; \boldsymbol{\theta})$,
where $m$ is the model of interest. One can also construct
regression models that are defined through the quantiles of the
conditional distribution of the response variable given the predictors \citep{Cade:2003}.
As the mean alone is not enough to characterize an entire distribution,
the classical regression method gives an incomplete picture of the
conditional distribution of the response variable given the
predictors \citep{Mosteller:1977}.

Quantile regression models are defined through the quantiles of the
conditional distribution of the response variable. Such models allow
one to evaluate relationships among variables through the conditional
median of the response variable, as well as the full range of other
conditional quantile functions. By supplementing the classical
regression model, which is defined at the conditional mean only,
quantile regression models provide a more complete
statistical analysis of the relationships among variables.
This is especially relevant when the range of variation in the dependent variable is of interest rather than its conditional mean.
For example, in a simple linear regression model with
heterogeneous variance linearly increasing/decreasing
relative to the predictor implies that
no single slope can characterize changes in
the conditional distribution of the response given
the predictor. In such a situation, by focusing
exclusively on changes in the conditional mean,
a classical linear regression model may
underestimate/overestimate the slopes
in the heterogeneous conditional distribution \citep{Cade:1999,Terrell:1996}.
Unequal variance in the conditional distribution implies that
there are many slopes describing the relationship between the
response and predictor. Quantile regression models can
estimate the multiple slopes and thus provide a more informative picture
of the relationships.

Let $m_{\tau}(x; \boldsymbol{\theta}_{\tau})$ be the $\tau$th quantile of the conditional
distribution of the response $y$ given a predictor $x$. Then
the piecewise linear quantile regression model with two breakpoints
is defined as following:

\begin{equation} \label{eqn:PLQR}
m_{\tau}(x_i; \boldsymbol{\theta}_{\tau}) = \left\{\begin{array}{lcr} \beta_{0\tau} + \beta_{1\tau} x_i & \text{for} & x_i \leq \alpha_{1\tau}\\
\beta_{0\tau} + \beta_{1\tau} x_i + \beta_{2\tau} (x_i - \alpha_{1\tau}) & \text{for} & \alpha_{1\tau} < x_i \leq \alpha_{2\tau}\\
\beta_{0\tau} + \beta_{1\tau} x_i + \beta_{2\tau} (x_i - \alpha_{1\tau}) + \beta_{3\tau} (x_i - \alpha_{2\tau}) & \text{for} & x_i > \alpha_{2\tau}\\
\end{array} \right.
\end{equation}
where $\alpha_{1\tau}$ and $\alpha_{2\tau}$ are the first and second
breakpoints defined at the $\tau$th quantile of the conditional distribution.

To estimate the parameters of the piecewise linear quantile regression model
the following objective function is minimized:
\[
\ \min_{\boldsymbol{\theta}_{\tau} \in R^p} \sum_{i=1}^n \rho_{\tau}\left(g_i(\boldsymbol{\theta}_{\tau})\right),
\]
where $g_i(\boldsymbol{\theta}_{\tau}) = (y_i - m_{\tau}(x_i; \boldsymbol{\theta}_{\tau}))$, 
assumed to be differentiable with respect to $\boldsymbol{\theta}_{\tau}$,
and $\rho_{\tau}$ is the
loss function defined as $\rho(u) = u\left(\tau - I(u < 0)\right)$,
for some $\tau \in (0, 1)$ and $I$ being the indicator function. Here,
$\boldsymbol{\theta}_{\tau}$ is the vector of parameters, with dimension $p = 6$,
for the regression model defined at the $\tau$th quantile. The above optimization
problem is solved using linear programming method \citep{Koenker:2005}. 
We used R package \texttt{quantreg} \citep{Quantreg:2018} to fit piecewise linear quantile regression
model where the initial values of the parameters are obtained from the
fitted piecewise linear regression model.

The advantage of quantile regression is that there is
no restriction for any distribution of the error term nor there is
any restriction in specifying the structure of the error variance.
Consequently, the quantile regression estimates
are well suited to constructing prediction intervals for the breakpoints
without the need to assume any parametric distribution for the estimates.

The piecewise linear quantile regression model, which is defined at conditional quantiles, provides much richer information 
in terms of estimating a relationship and breakpoints than the piecewise linear
regression model, which is defined at the conditional mean. The collection of estimates of a breakpoint
identified from different quantile regression curves provide an appropriate confidence interval for the
index breakpoint. Moreover, the upper and lower quantile regression curves provide
a prediction band for the data points.

\subsubsection{Calculation of area within the prediction band with rationale}

We calculated the area within the confidence band by introducing grids in
the horizontal axis. The smaller the grid cells the better the efficiency
in calculation. Within each small grid cell, we calculated the area of the
rectangle (height $\times$ width) which is sandwiched between
the lower and upper limits of the band. The
areas of the rectangles in the grid that fell within the prediction bands of the relationship were then added together to provide an estimate for the area of the
prediction band.

The calculation of area within the band allows us to compare the relative precision of the prediction bands
obtained from different methods. In this paper, we compare $80\%$
prediction band as it is available for the $3$ methods we 
have introduced.

\section{Applications} \label{sect:results}

\subsection{Relating a wetland fish index of biotic integrity to variation in agricultural stress among wetlands}

The index of biotic integrity (IBI)
provides a measure of the ecological condition of an ecoregion or watershed.
\cite{Uzarski:2005} and \cite{Bhagat:2007}
validated a multimetric index of biotic integrity by assessing
fish community composition in stands of bulrush (\textit{Schoenoplectus, spp})
in $30$ coastal wetlands distributed across the 
US coast of the Great Lakes (Top of Table \ref{tab:IBI:AgSt:meanCI}).
Numerous researchers
have quantified agricultural and land use as the major cause of natural habitat alteration, 
affecting fish communities in lakes and rivers \citep{Brazner:1997,Crosbie:1999}.
\cite{Danz:2005}
derived an index summarizing agricultural stress to characterize disturbance
to natural land using GIS based data across the US Great Lakes coastline.
In this application, we regressed the fish index of biotic integrity against
the agricultural stress gradient \citep{Bhagat:2007}. Theoretically, the fish IBI
scores vary from $0$ to $100$, with larger scores representing better ecological
health of the fish community sampled at a coastal shoreline draining a watershed. The
agricultural stress gradient, a scaled principal component analysis (PCA) score
was rescaled from its original values \citep{Danz:2005} to a 0 to 1 range
with larger numbers reflecting more extensive agricultural activities.
The original PCA scores ranged from  $-1.5634$ to $+1.6733$,
and the adjusted PCA scores 
were rescaled according to the formula:
\[
\ \text{Scaled agricultural stress} = 0.341774 +  (0.196037 \times \text{Reported agricultural stress score}).
\]
The overall mean of fish IBI is $44.40$ with $95\%$ confidence interval of $39.47$ to $49.33$.
The overall mean of agricultural stress is $0.32$ with $95\%$ confidence interval of $0.27$ to $0.38$.
\cite{Bhagat:2007} concluded that the fish IBI scores exhibited a negative linear correlation with
respect to agricultural stress, but suggested the presence of threshold responses;
however, they did not quantitatively test for the presence of breakpoints.

\subsection{Relating cyanobacteria biomass to total phosphorus concentrations among lakes}

Increasing prevalence of cyanobacteria blooms and their biomass is a global phenomenon \citep{Bullerjahn:2016}.
Cyanobacteria blooms are manifestations of eutrophication and are directly related to risks to human and
animal health \citep{Downing:2001,Sondergaard:2018}. Cyanobacteria-dominated blooms generate compounds
that can be acutely toxic \citep{Campos:2010,Roegner:2014}, and that are linked to diseases such as carcinoma
\citep{Labine:2009,Lone:2015}. Mitigation of these risks requires understanding causes of blooms in lakes and rivers.
Biomass is a standard index of the extent and concentration of cyanobacteria and often used as a proxy
for the risk of toxicity of harmful algal blooms.

The frequency of cyanobacterial blooms is rising across Canada \citep{Pick:2016} as well as elsewhere.
\cite{Beaulieu:2014} used data from 
$149$ lakes collected across three regions in Canada 
and fitted a linear
regression model to predict cyanobacterial biomass ($\mu$g/L) from total phosphorus concentrations (TP, $\mu$g/L).

Because the relationships between $log_{10}(\text{cyanobacterial biomass})$ and $log_{10}(\text{total phosphorus})$ has been argued to be better
represented as a sigmoidal function with abrupt changes evident at two positions defined by log(total
phosphorus) concentration, we used the piecewise linear regression model to estimate the breakpoints
and generated confidence bands for the response variable given the predictor.

The data presented in \cite{Beaulieu:2014} were collected from $43$ sites from Alberta, $10$ from British Columbia,
and $97$ from Ontario (Bottom of Table \ref{tab:IBI:AgSt:meanCI}).
The data were provided by the Ministries of the Environment of Alberta, British Columbia and Ontario.

The overall mean and $95\%$ confidence interval of the $log_{10}(\text{cyanobacterial biomass})$ is $2.12$ and ($1.96-2.28$),
respectively.
The overall mean of $log_{10}(\text{total phosphorus})$ is $1.20$ with $95\%$ confidence interval $(1.13-1.27)$.
The mean values are greatest in Alberta followed by British Columbia and Ontario. 
The total phosphorus concentration is
significantly higher in Alberta lakes than in those from British Columbia and Ontario. The mean of cyanobacteria biomass and total phosphorus concentration in British Columbia and Ontario are not significantly different.

\begin{table}[ht]
\centering
\caption{\label{tab:IBI:AgSt:meanCI}Means and $95\%$ confidence intervals of means of the two variables for the fish index of biotic integrity score and agricultural stress data (top) and for the
	cyanobacteria biomass and total phosphorus data (bottom).}
\begin{tabular}{||l|lcr|rr||}
  \hline\hline
  \multicolumn{6}{||c||}{Fish IBI score and agricultural stress}\\\hline
\multirow{2}{*}{Source} & \multirow{2}{*}{Variables} & \multirow{2}{*}{$n$} & \multirow{2}{*}{Mean} & \multicolumn{2}{c||}{$95\%$ CI of mean}\\\cline{5-6}
& & & & lower & upper\\\hline
	\multirow{2}{*}{GLEI\footnotemark[1]} & Fish IBI & \multirow{2}{*}{$13$} & 42.23 & 36.57 & 47.90\\
	& Agg. stress & & 0.30 & 0.20 & 0.40\\\hline
	\multirow{2}{*}{Uzarski\footnotemark[2]} & Fish IBI & \multirow{2}{*}{$17$} & 46.06 & 38.04 & 54.08\\
	& Agg. stress & & 0.34 & 0.27 & 0.40\\\hline
	\multirow{2}{*}{Combined} & Fish IBI & \multirow{2}{*}{$30$} & 44.40 & 39.47 & 49.33\\
	& Agg. stress & & 0.32 & 0.27 & 0.38\\\hline\hline
  \multicolumn{6}{||c||}{Cyanobacteria biomass and total phosphorus}\\\hline
\multirow{2}{*}{Source} & \multirow{2}{*}{Variables} & \multirow{2}{*}{$n$} & \multirow{2}{*}{Mean} & \multicolumn{2}{c||}{$95\%$ CI of mean}\\\cline{5-6}
& & & & lower & upper\\\hline
			\multirow{2}{*}{Alberta} & Cyan & \multirow{2}{*}{$43$} & 2.86 & 2.62 & 3.10\\
			& TP & & 1.54 & 1.44 & 1.63\\\hline
			\multirow{2}{*}{BC} & Cyan & \multirow{2}{*}{$10$} & 2.40 & 1.74 & 3.06\\
			& TP & & 1.12 & 0.81 & 1.42\\\hline
			\multirow{2}{*}{Ontario} & Cyan & \multirow{2}{*}{$97$} & 1.76 & 1.58 & 1.94\\
			& TP & & 1.06 & 0.98 & 1.14\\\hline
			\multirow{2}{*}{Combined} & Cyan & \multirow{2}{*}{$150$} & 2.12 & 1.96 & 2.28\\
			& TP & & 1.20 & 1.13 & 1.27\\\hline\hline
	\end{tabular} 
\end{table}
\footnotetext[1]{Great Lakes Environmental Indicators (GLEI) by \cite{Bhagat:2007}}
\footnotetext[2]{\cite{Uzarski:2005}}

\section{Results}
\subsection{Relationships between fish index of biotic integrity and agricultural stress}

\subsubsection{Loess and bootstarap confidence band} \label{sect:loess:IBIvsAgSt}

\begin{figure}[ht]
\begin{center}
\includegraphics[width=0.77\textwidth]{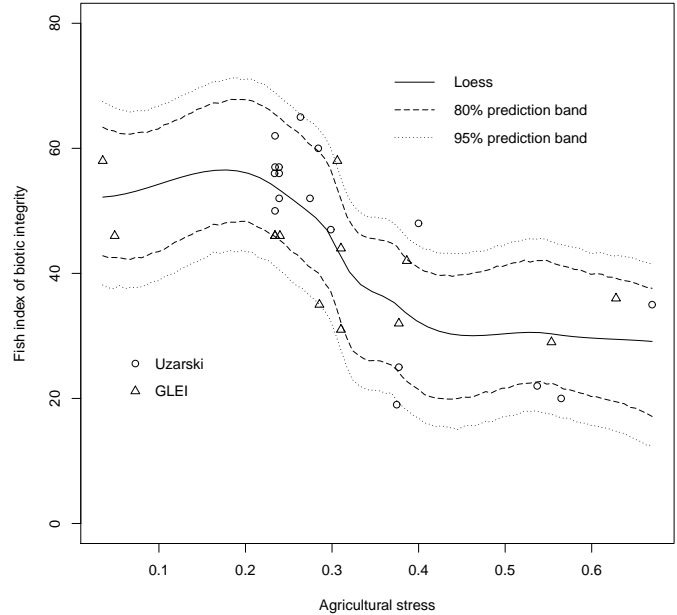}
\caption{\label{Loess_CB_Bootstrap}The scatter plot of
fish index of biotic integrity against agricultural stress
highlighting labels for Uzarski and GLEI sites. The scatter
plot displays loess along with $80\%$ and $95\%$
bootstrap prediction bands, which cover surface areas of $12.479$ and $17.666$ square units, respectively.}
\end{center}
\end{figure}

The relationship between fish IBI and agricultural stress was negative (Figure \ref{Loess_CB_Bootstrap}): the fish IBI decreased with the increase of agricultural stress.
The nonparametric loess suggests the possible presence of two breakpoints -
one around agricultural stress value of $0.22$ and the other around $0.45$.
The Fish IBI score was independent of agricultural stress for agricultural stress
scores up to $0.22$, decreased sharply between stress values of $0.22$
to $0.45$, and reached its minimum at stress values greater than $0.45$.
The $80\%$ and $95\%$ bootstrap prediction bands for the fish IBI were
obtained from $10,000$ bootstrap samples. The loess and bootstrapped prediction bands provide
an idea about possible locations and spread of future data points.
The $80\%$ and $95\%$
bootstrapped prediction bands covered surface areas of $12.479$ and $17.666$ square units, respectively.

\subsubsection{Piecewise linear regression model} \label{sect:PLR:IBIvsAgSt}

Piecewise linear regression fitted to the fish IBI and agricultural stress
data identified two breakpoints at agricultural stress scores of $0.263$ and $0.488$,
respectively (Table \ref{CIPLRM}), broadly corresponding to the location of inflection points
identified by the loess model.
Table \ref{CIPLRM} also shows the estimates, standard errors, t-values, p-values and $95\%$ confidence intervals
of the breakpoints and slopes of the piecewise linear regression model
fitted to the fish IBI versus agricultural stress data. The $95\%$ confidence
intervals for the first and second breakpoints $\alpha_1$ and $\alpha_2$
are $(0.196, 0.331)$ and $(0.391, 0.585)$, respectively.
There is no overlap between the confidence intervals for the two estimated breakpoints.
Thus, the two breakpoints are significantly different from one other.

The $95\%$ confidence interval of the regression coefficient ($\beta_1$) for the first
segment of the regression model is $(-70.110, 78.920)$.
As the confidence interval contains $\beta_1 = 0$ inside with $t$ and $p$ values of $0.122$ and $0.904$, respectively,
the relationship between fish IBI and agricultural stress is statistically
nonsignificant in this segment. That is, Fish IBI did not change as Agricultural
Stress changed from $0$ to $0.263$. The $95\%$ confidence interval of the
slope $(\beta_1 + \beta_2)$ in the second segment of the regression model is $(-272.800,-50.670)$.
As the confidence interval does not contain $\beta_1 + \beta_2 = 0$, with $t$ and $p$ values of $-3.005$ and $0.006$, respectively, the 
relationship between fish IBI and agricultural stress is statistically significant in this segment.
That is, fish IBI decreases significantly with the increase of agricultural stress between
$0.263$ to $0.488$.
The $95\%$ confidence interval of slope $(\beta_1 + \beta_2 + \beta_3)$ in the third segment of the regression
model is $(-54.240,278.800)$. As the confidence interval contains $\beta_1 + \beta_2 + \beta_3 = 0$
inside with $t$ and $p$ values of $1.392$ and $0.177$, respectively, the relationship between fish IBI and agricultural stress is statistically insignificant in this segment.
That is, fish IBI does not reflect the changes as agricultural stress changes from $0.488$ and $1.00$.

\begin{table}[ht]
\centering
\caption{\label{CIPLRM}Estimates, standard errors (SE), t-values ($t$), p-values ($p$) and $95\%$ confidence intervals ($95\%$ CI)
of the breakpoints (breaks) and slopes of the piecewise linear regression lines 
fitted to the fish index of biotic integrity (IBI) score versus agricultural stress data. The statistically significant slopes are highlighted by light grey color.}
\begin{tabular}{||ll|rrccrr||}
  \hline\hline
	\multirow{3}{*}{Names} & \multirow{3}{*}{Parameters} & \multicolumn{6}{c||}{Fish IBI score and agricultural stress}\\\cline{3-8}
	& & \multirow{2}{*}{Estimates} & \multirow{2}{*}{SE} & \multirow{2}{*}{$t$} & \multirow{2}{*}{$p$} & \multicolumn{2}{c||}{$95\%$ CI}\\\cline{7-8}
	& & & & &&Lower & Upper\\\hline
	\multirow{2}{*}{Breaks} & $\alpha_1$ & 0.263 & 0.033 & 7.970 & 0.000 & 0.196 & 0.331\\
	& $\alpha_2$ & 0.488 & 0.047 & 10.383 & 0.000 & 0.391 & 0.585\\\hline
	\multirow{4}{*}{Slopes} & $\beta_1$ & 4.405 & 36.100 & 0.122 & 0.904 & -70.110 & 78.920\\
	& $\beta_1 + \beta_2$ & \cellcolor{gray!37}-161.700 & 53.810 & -3.005 & $0.006^{*}$ & -272.800 & -50.670\\
& $\beta_1 + \beta_2 + \beta_3$ & 112.300 & 80.680 & 1.392 & 0.177 & -54.240 & 278.800\\
   \hline\hline
\end{tabular}
\end{table}

\begin{figure}[ht]
\begin{center}
\includegraphics[width=0.77\textwidth]{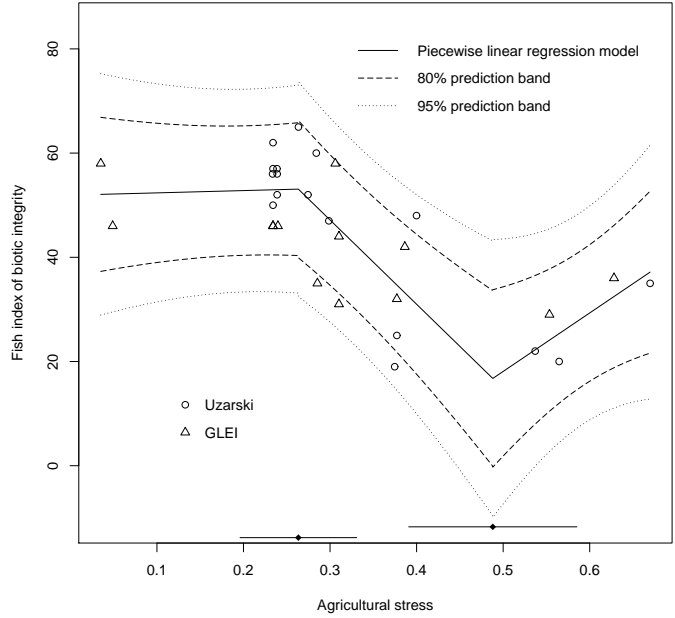}
\caption{\label{PLRM_Pred_Int}The scatter plot of
fish index of biotic integrity against agricultural stress
highlighting labels for Uzarski and GLEI sites. The scatter
plot displays the fitted piecewise linear regression model along with $80\%$ and $95\%$
prediction bands which cover surface areas of $17.240$ and $27.000$ square units, respectively. The two solid lines
along the horizontal axis are the marginal $95\%$ confidence intervals of the breakpoints.}
\end{center}
\end{figure}

Figure \ref{PLRM_Pred_Int} shows the piecewise linear regression model
fitted to the fish IBI and agricultural stress data. This figure depicts
the two breakpoints $0.263$ and $0.488$ along with the $95\%$
confidence intervals (see the two solid lines along the horizontal axis).
This plot also shows the $80\%$ and $95\%$ prediction bands
for the response variable fish IBI given the predictor agricultural stress
which provide an idea about the positions
of the future data points. 
The $80\%$ and $95\%$ prediction bands cover surface areas of $17.240$
and $27.000$ square units, respectively.
The prediction bands are wider in this figure,
in contrast to Figure \ref{Loess_CB_Bootstrap},
as the model contains $6$ parameters and leaving only $30$ data points
for estimation.

\subsubsection{Piecewise linear quantile regression} 

\label{sect:PLQR:IBIvsAgSt}
\begin{figure}[ht]
\centering
\includegraphics[width=0.77\textwidth]{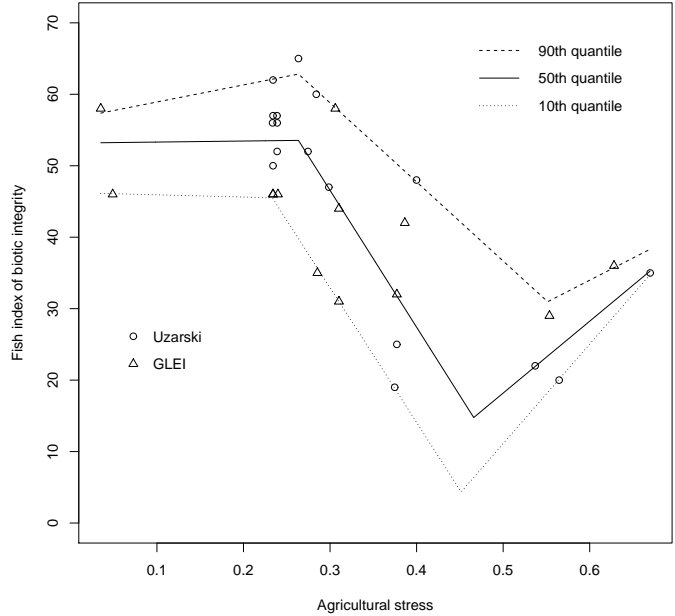}
\caption{Piecewise linear quantile regression models for quantiles $\tau = 0.10$, $0.50$, and $0.90$ applied to the fish index of biotic integrity and agricultural stress data. The models with quantiles $0.10$ and $0.90$ provide $80\%$ prediction band for the response given the predictor. The $80\%$ prediction band covers surface area of $12.438$ square units.}
\label{fig:QRM_IBI}
\end{figure}

A piecewise linear quantile regression model provides much richer
information in terms of estimating relationship and breakpoints
than the piecewise linear regression model which goes through the means
of each segment.
The collection of breakpoints from different quantile regression
curves provides confidence intervals for the breakpoints (Table \ref{tab:breaks:IBI}).
Moreover, the upper and lower quantile regression curves
provide 
narrower prediction bands ($12.438$ square units only)
for the data points (Figure \ref{fig:QRM_IBI})
than the other two models.
The quantile regression curves at the $10$th and $90$th
quantiles closely tracked the lower and upper bounds of the data points, respectively.

The fitted quantile regression curves each provide paired estimates for the values of first and second breakpoints on the stress
axis (Table \ref{tab:breaks:IBI}). The smallest and the largest
breakpoints are highlighted by light and dark gray, respectively.
The collection of estimates for the first breakpoint ranges from $0.233$ to $0.284$ and can be
considered as the $80\%$ confidence interval for $\alpha_1$. The collection of estimates for the second
breakpoint ranges from $0.448$ and $0.564$ and can be considered as the $80\%$ confidence interval
for $\alpha_2$. So the first breakpoint occurs within the interval $0.233$ to $0.284$ and the second
breakpoint is situated somewhere between $0.448$ and $0.564$.

\begin{table}[ht]
\centering
\caption{\label{tab:breaks:IBI}Estimates of the breakpoints $\alpha_1$ and $\alpha_2$ of the piecewise linear quantile regression models for quantiles $\tau = (0.10, 0.20, 0.30, 0.40, 0.50, 0.60, 0.70, 0.80, 0.90)$ applied to the fish IBI score and agricultural stress data. The grid of quantiles covers $(0.90 - 0.10)\times 100 = 80\%$
of the total space. Hence, the
smallest and largest breakpoints - which are highlighted by light and dark grey, respectively - provide $80\%$ confidence intervals for the breakpoints.}
\begin{tabular}{||c|cc|c|cc||}
  \hline\hline
Quantiles & \multicolumn{2}{c|}{Breakpoints} & Quantiles & \multicolumn{2}{c||}{Breakpoints}\\\cline{2-3}\cline{5-6}
$\tau$ & $\alpha_1(\tau)$ & $\alpha_2(\tau)$ & $\tau$ & $\alpha_1(\tau)$ & $\alpha_2(\tau)$\\\hline
0.10 & \cellcolor{gray!27}0.233 & 0.452 & 0.60 & 0.255 & 0.476\\
0.20 & 0.239 & 0.450 & 0.70 & 0.264 & 0.533\\
0.30 & 0.233 & 0.450 & 0.80 & \cellcolor{gray!57}0.284 & \cellcolor{gray!57}0.564\\
0.40 & 0.270 & \cellcolor{gray!27}0.448 & 0.90 & 0.264 & 0.552\\
0.50 & 0.264 & 0.466 & - & - & - \\ 
   \hline\hline
\end{tabular}
\end{table}

\subsection{Relationship between cyanobacteria biomass and total phosphorus}

Subsections \ref{sect:loess:CyanvsTP}, \ref{sect:PLR:CyanvsTP}, and \ref{sect:PLQR:CyanvsTP}
contain results from statistical
methods applied to the cyanobacteria biomass and total phosphorus data.

\subsubsection{Loess and bootstrap prediction band} \label{sect:loess:CyanvsTP}

\begin{figure}[ht]
	\begin{center}
		\includegraphics[width=0.77\textwidth]{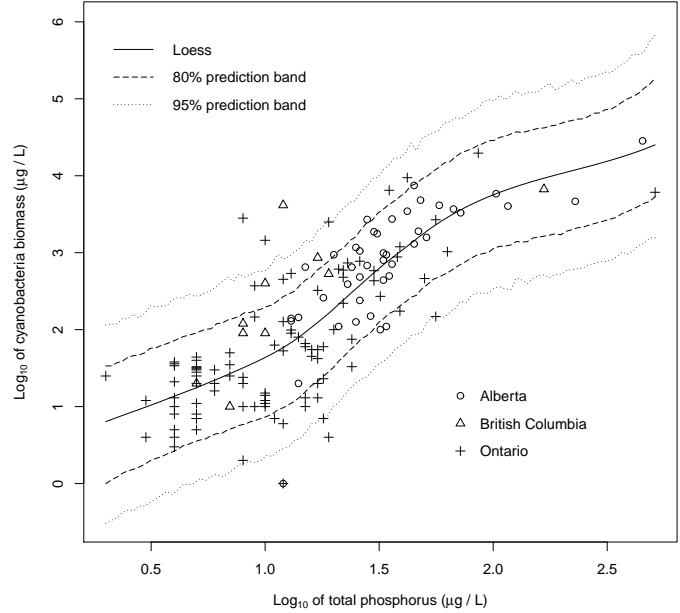}
		\caption{\label{Loess_PB_Bootstrap_Cyan}The scatterplot of
			log of cyanobacteria biomass against log of total phosphorus
			highlighting labels for the lakes in Alberta, British Columbia and Ontario. The scatter
			plot displays loess along with $80\%$ and $95\%$
			bootstrap prediction bands, which cover surface areas of $3.456$ and $5.936$ square units, respectively.}
	\end{center}
\end{figure}

The fitted model loess and its $80\%$ and $95\%$ prediction
bands indicated that there was a positive relationship between
$log_{10}(\text{cyanobacteria biomass})$ and $log_{10}(\text{total phosphorus})$ (Figure \ref{Loess_PB_Bootstrap_Cyan}).
Discontinuities in the trend line suggested the potential of
finding two candidate breakpoints one at $log(\text{TP})$ of approximately $1.20$
($15.85$ $\mu$g/L)
and the other at around $log(\text{TP})$ of $1.70$
($50.12$ $\mu$g/L).
Cyanobacteria biomass increased steadily as function of $log(\text{TP})$ up to a
value of $1.20$, rises sharply between $log(\text{TP})$ values between $1.20$ and $1.70$,
and then rises more slowly at greater $log(\text{TP})$ concentrations. The $80\%$ and $95\%$
bootstrap prediction bands (generated from $10000$ bootstrap samples)
identify the potential range of cyanobacteria biomass of a given value of $log(\text{TP})$.
The $80\%$ and $95\%$ prediction bands cover surface areas of $3.456$ and $5.936$
square units, respectively.

\subsubsection{Piecewise linear regression model} \label{sect:PLR:CyanvsTP}

The piecewise linear regression model (equation \ref{eqn:PLRM})
fitted to the cyanobacteria biomass 
identified lower and upper breakpoints ($\alpha_1$ and $\alpha_2$)
at $log(\text{TP})$ of 
$1.212$ ($16.293$ $\mu$g/L)
and $1.624$ ($42.073$ $\mu$g/L),
respectively (Table \ref{CIPLRM_Cyan}).

Table \ref{CIPLRM_Cyan} also shows the estimates, standard errors, t-values, p-values
and $95\%$ confidence intervals of the breakpoints and slopes
of the piecewise linear regression model fitted to the cyanobacteria biomass
vs total phosphorus data.
The statistically significant slopes are highlighted by light gray shading.
The $95\%$ confidence intervals for the breakpoints $\alpha_1$ and $\alpha_2$
are $(1.026, 1.399)$ and $(1.418, 1.830)$, respectively. The two breakpoints
are statistically different as there is no overlap between the two confidence intervals.

\begin{figure}[ht]
\begin{center}
\includegraphics[width=0.77\textwidth]{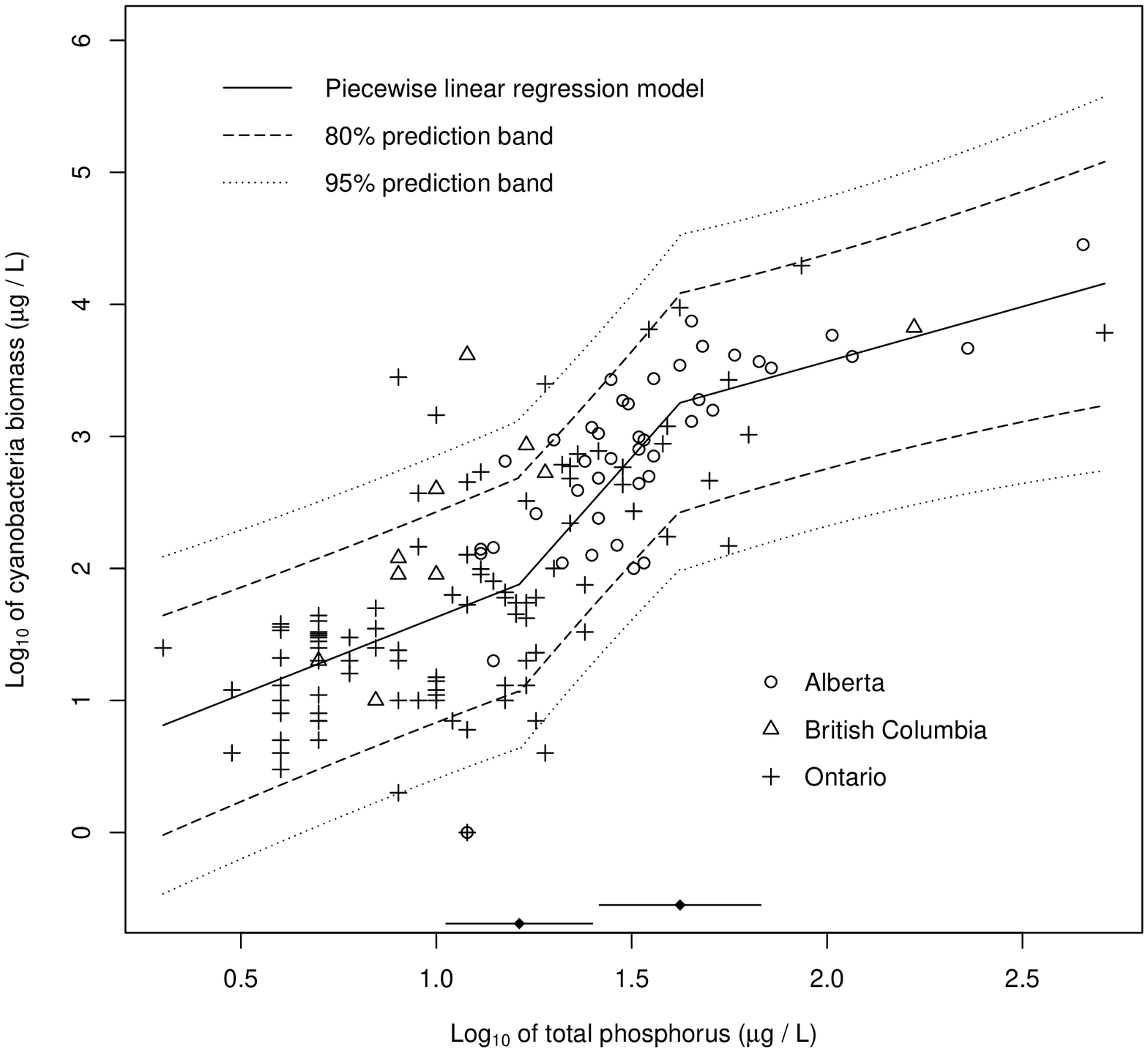}
\caption{\label{PLRM_Pred_Int_Cyan}The scatterplot of
log of cyanobacteria biomass against log of total phosphorus
highlighting the labels for the lakes in Alberta, British Columbia and Ontario. The scatter
plot displays the fitted piecewise linear regression model along with $80\%$ and $95\%$
prediction bands, which cover surface areas of $3.956$ and $6.073$ square units, respectively. The two solid lines
along the horizontal axis are the marginal $95\%$ confidence intervals of the breakpoints.}
\end{center}
\end{figure}

The values and trends of the three slopes estimated by the piecewise regression
analysis (Table \ref{CIPLRM_Cyan}; Figure \ref{PLRM_Pred_Int_Cyan})
were consistent in relative magnitude with the patterns subjectively described by the loess model.
Values for regression slope coefficients in the first and second segments of the regression
lines were both significantly greater than zero $(b_1 = 1.172, t = 3.595, p = 0.000; b_1 + b_2 = 3.344, t = 4.791, p = 0.000)$.
Although the estimated slope of the third segment
of the regression line was greater than zero, the value was not quite
statistically significant $(b_1 + b_2 + b_3 = 0.831, t = 1.884, p = 0.062)$.
Figure \ref{PLRM_Pred_Int_Cyan} displays $80\%$ and $95\%$
prediction bands, which cover surface areas of $3.956$
and $6.073$ square units, respectively.
The prediction bands for the data points are compact
as there are many data points $(150, \text{here})$ in this application.
The prediction bands provide an idea about the possible range of values of log
of cyanobacteria biomass for a given value of log of total phosphorus.

\begin{table}[ht]
\centering
\caption{\label{CIPLRM_Cyan}Estimates, standard errors (SE), t-value $(t)$, p-value $(p)$ and $95\%$ confidence intervals ($95\%$ CI) of the breakpoints (breaks) and slopes of the piecewise linear regression model fitted to the cyanobacteria biomass and total phosphorus data. The statistically significant slopes are highlighted by light grey color.}
\begin{tabular}{||ll|rrccrr||}
  \hline\hline
	\multirow{3}{*}{Names} & \multirow{3}{*}{Parameters} & \multicolumn{6}{c||}{Cyanobacteria biomass vs total phosphorus}\\\cline{3-8}
	& & \multirow{2}{*}{Estimates} & \multirow{2}{*}{SE} & \multirow{2}{*}{$t$} & \multirow{2}{*}{$p$} &\multicolumn{2}{c||}{$95\%$ CI}\\\cline{7-8}
	& & & & &&Lower & Upper\\\hline
	\multirow{2}{*}{Breaks} & $\alpha_1$ & 1.212 & 0.094 & 12.894 & 0.000 & 1.026 & 1.399\\
	& $\alpha_2$ & 1.624 & 0.104 & 15.615 & 0.000 & 1.418 & 1.830\\\hline
	\multirow{4}{*}{Slopes} & $\beta_1$ & \cellcolor{gray!37}1.172 & 0.326 & 3.595 & $0.000^{*}$ & 0.528 & 1.815\\
	& $\beta_1 + \beta_2$ & \cellcolor{gray!37}3.344 & 0.698 & 4.791 & $0.000^{*}$ & 1.964 & 4.725\\
& $\beta_1 + \beta_2 + \beta_3$ & 0.831 & 0.441 & 1.884 & 0.062 & -0.041 & 1.702\\
   \hline\hline
\end{tabular}
\end{table}

\subsubsection{Piecewise linear quantile regression} \label{sect:PLQR:CyanvsTP}

\begin{figure}[ht]
\centering
\includegraphics[width=0.77\textwidth]{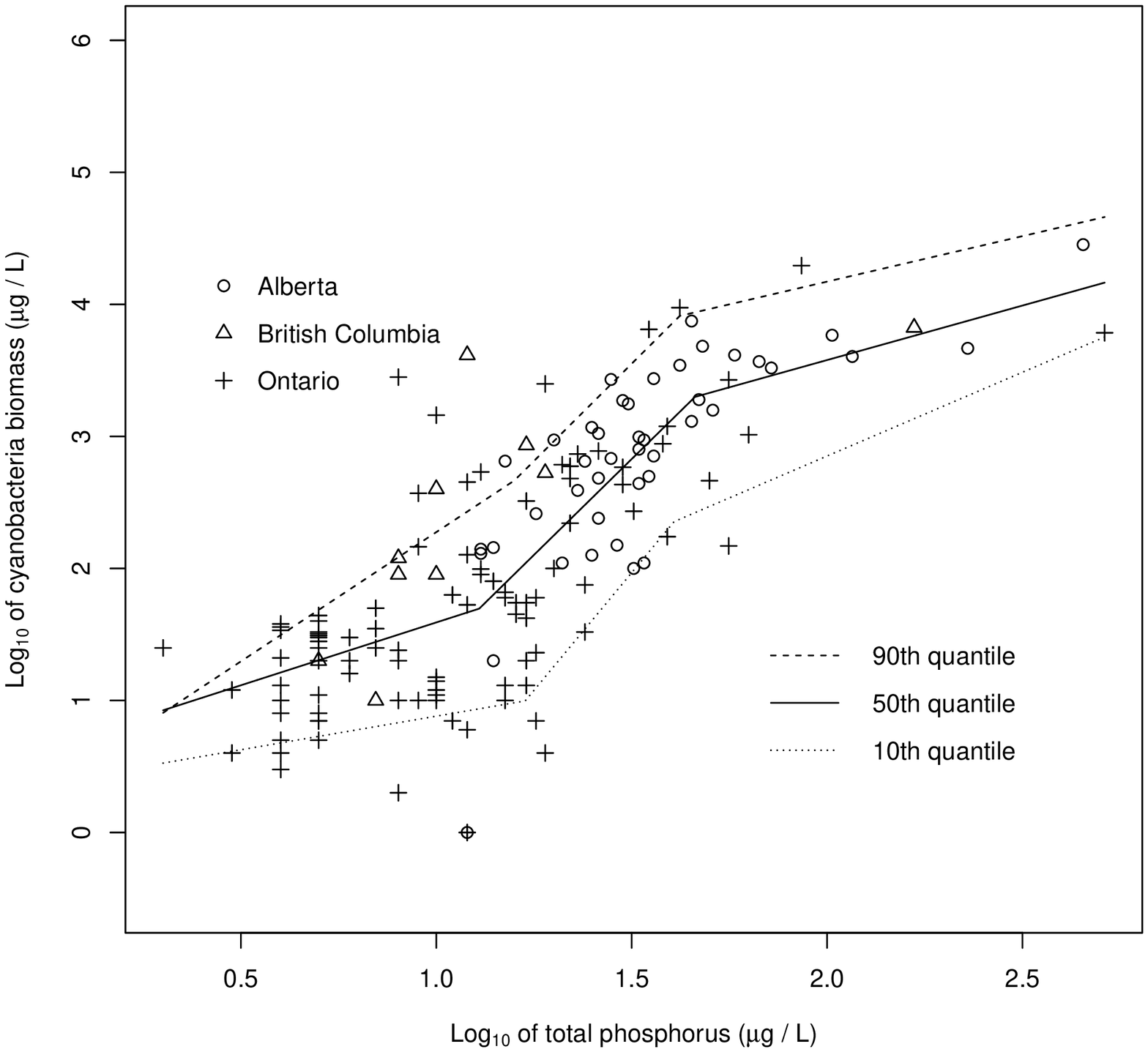}
\caption{\label{fig:QRM_Cyan}Piecewise linear quantile regression models for quantiles $\tau = 0.10$, $0.50$, and $0.90$ applied to
the log of cyanobacteria biomass and log of total phosphorus data. The models with quantiles $0.10$ and $0.90$ provide $80\%$ prediction band for the data points. The $80\%$ prediction band covers $2.951$ square units of the surface area.}
\end{figure}

Figure \ref{fig:QRM_Cyan} shows the fitted piecewise linear quantile regression model (equation \ref{eqn:PLQR})
for conditional quantiles $\tau = 0.10$, $0.50$, and $0.90$ bounding the log of cyanobacteria biomass and log of
total phosphorus
relationship. The quantile regression curves at $10$th and $90$th quantiles
provide a prediction band for the range of the cyanobacterial biomass values expected for a given
total phosphorus concentration.
The $80\%$ prediction band covers $2.951$ square units of the surface area.

The quantile regression models generate an estimate of the location of the first and second
break points, for each of the quantile boundaries (Table \ref{tab:breaks:Cyan}).
The estimated values for the first break points $\alpha_1$ ranged from $1.066$ ($11.641$ $\mu$g/L)
to $1.230$ ($16.982$ $\mu$g/L).
The values for the second break point $\alpha_2$ ranged from the $1.585$ ($38.459$ $\mu$g/L)
to $1.662$ ($45.920$ $\mu$g/L). The $80\%$ prediction intervals for $\alpha_1$ and $\alpha_2$
are $(1.066, 1.230)$ and $(1.585, 1.662)$, respectively.

\begin{table}[ht]
\centering
\caption{\label{tab:breaks:Cyan}Estimates of the breakpoints $\alpha_1$ and $\alpha_2$ of the piecewise linear quantile regression models for quantiles $\tau = 0.10$, $0.20$, $0.30$, $0.40$, $0.50$, $0.60$, $0.70$, $0.80$ and $0.90$ applied to the log of cyanobacteria biomass and log of total phosphorus data. 
The grid of quantiles covers $(0.90 - 0.10)\times 100 = 80\%$
of the total space.
Hence, the
smallest and largest breakpoints - which are highlighted by light and dark grey, respectively - provide $80\%$ confidence intervals for the breakpoints.}
\begin{tabular}{||c|cc|c|cc||}
  \hline\hline
Quantiles & \multicolumn{2}{c|}{Breakpoints} & Quantiles & \multicolumn{2}{c||}{Breakpoints}\\\cline{2-3}\cline{5-6}
$\tau$ & $\alpha_1(\tau)$ & $\alpha_2(\tau)$ & $\tau$ & $\alpha_1(\tau)$ & $\alpha_2(\tau)$\\\hline
0.10 & 1.228 & 1.609 & 0.60 & 1.176 & 1.641\\
0.20 & 1.201 & \cellcolor{gray!27}1.585 & 0.70 & 1.155 & 1.623\\
0.30 & \cellcolor{gray!27}1.066 & 1.645 & 0.80 & 1.146 & 1.598\\
0.40 & \cellcolor{gray!57}1.230 & 1.646 & 0.90 & 1.200 & 1.623\\
0.50 & 1.110 & \cellcolor{gray!57}1.662 & - & - & - \\ 
   \hline\hline
\end{tabular}
\end{table}

\section{Comparison of confidence intervals and prediction bands} \label{sect:ci:pb}

We examined which method provided the narrowest confidence intervals of
the breakpoints. We fixed the confidence coefficient at $0.80$
as the piecewise linear quantile regression model provides $80\%$ confidence intervals
for the breakpoints. As loess does not provide estimates for the breakpoints,
we could not calculate the width for loess. However, one can use loess to subjectively estimate the
positions of the breakpoints and their intervals.

Table \ref{tab:comp:ci} compares the width of the $80\%$ confidence intervals
of the breakpoints $\alpha_1$ and $\alpha_2$ for the piecewise linear
regression model (PLRM) and piecewise linear quantile regression model (PQRM).
The first part of this table shows results for the fish 
index of biotic integrity and agricultural stress data. 
The widths of the first breakpoints $\alpha_1$
are $0.306 - 0.220 = 0.086$ and $0.284 - 0.233 = 0.051$
for PLRM and PQRM, respectively. 
For $\alpha_1$, the PQRM generated a substantially narrower interval.
Similarly, the widths of the $80\%$ confidence intervals for the second
breakpoint $\alpha_2$ are $0.124$ and $0.116$ for PLRM and PQRM, respectively.
Again, the PQRM generated a narrower interval.

The second part of Table \ref{tab:comp:ci} summarizes the cyanobacteria biomass
and total phosphorus data. For the first breakpoint $\alpha_1$, the widths of the
$80\%$ confidence intervals for PLRM and PQRM are $0.243$ and $0.164$, respectively.
For the second breakpoint $\alpha_2$, the widths of the
$80\%$ confidence intervals for PLRM and PQRM are $0.269$ and $0.077$, respectively.
Again, for a given confidence coefficient $0.80$ the smaller width is produced
by PQRM. Thus, we advocate that the piecewise linear quantile regression model
provides more precise confidence intervals for the breakpoints.

\begin{table}[ht]
\centering
\caption{\label{tab:comp:ci}Comparison of width of the $80\%$ confidence intervals
of the breakpoints $(\alpha_1$ and $\alpha_2)$ for the piecewise linear regression model (PLRM)
and piecewise linear quantile regression model (PQRM). The smaller the width, the better
the method. The top portion belongs
to the fish IBI score and agricultural stress data and the bottom portion belongs to the
cyanobacteria biomass and total phosphorus data. The smaller width in each row, which is
highlighted by the dark grey color, belongs to the piecewise linear quantile regression model.}
\begin{tabular}{||c|ccc|ccc||}
  \hline\hline
Methods & \multicolumn{3}{c|}{PLRM} & \multicolumn{3}{c||}{PQRM}\\\hline
\multicolumn{7}{||c||}{Fish IBI score vs agricultural stress}\\\hline
\multirow{2}{*}{Breakpoints} & \multicolumn{3}{c|}{$80\%$ confidence intervals} & \multicolumn{3}{c||}{$80\%$ confidence intervals}\\
& Lower & Upper & Width & Lower & Upper & Width\\\hline
$\alpha_1$ & 0.220 & 0.306 & 0.086 & 0.233 & 0.284 & \cellcolor{gray!57}0.051 \\ 
$\alpha_2$ & 0.426 & 0.550 & 0.124 & 0.448 & 0.564 & \cellcolor{gray!57}0.116 \\\hline 
\multicolumn{7}{||c||}{Cyanobacteria biomass vs total phosphorus}\\\hline
\multirow{2}{*}{Breakpoints} & \multicolumn{3}{c|}{$80\%$ confidence intervals} & \multicolumn{3}{c||}{$80\%$ confidence intervals}\\
& Lower & Upper & Width & Lower & Upper & Width\\\hline
$\alpha_1$ & 1.091 & 1.334 & 0.243 & 1.066 & 1.230 & \cellcolor{gray!57}0.164 \\ 
$\alpha_2$ & 1.489 & 1.758 & 0.269 & 1.585 & 1.662 & \cellcolor{gray!57}0.077 \\ 
   \hline\hline
\end{tabular}
\end{table}

We also compared the surface areas bounded by the prediction band for each of
the three methods.
For a given confidence coefficient, the method covering the smallest
surface area is considered the best. Table \ref{tab:comp:predint}
compares the surface areas in square units of the $80\%$ prediction bands
for bootstrapped loess (BL), piecewise linear regression model (PLRM)
and piecewise linear quantile regression model (PQRM). The top part of Table \ref{tab:comp:predint}
shows results for the fish IBI and agricultural stress data. For the
confidence coefficient of $0.80$, the surface areas covered by BL, PLRM and
PQRM are $12.479$, $17.240$, and $12.438$ square units, respectively.
The smallest surface area $12.438$, highlighted by dark grey
color, was derived from the PQRM.

The bottom part of Table \ref{tab:comp:predint} shows results for
the cyanobacteria biomass and total phosphorus data. The surface areas
of $80\%$ prediction bands are $3.456$, $3.956$, and $2.951$ square
units, respectively. 
Even though bootstrapped loess
provides surface area which is close to piecewise linear quantile regression
model, we prefer the latter method as it provides better
estimates of the breakpoints.

\begin{table}[ht]
\centering
\caption{\label{tab:comp:predint}Comparison of the surface areas in square units of the $80\%$ prediction bands
for bootstrapped loess (BL), piecewise linear regression model (PLRM)
and piecewise linear quantile regression model (PQRM). The smaller the surface area, the better
the method. The top portion belongs
to the fish IBI score and agricultural stress data and the bottom portion belongs to the
cyanobacteria biomass and total phosphorus data. The smaller surface area in each row, which is
highlighted by the dark grey, belongs to the piecewise linear quantile regression model.}
\begin{tabular}{||l|ccc||}
  \hline\hline
\multicolumn{4}{||c||}{Fish IBI score vs agricultural stress}\\\hline
Methods & BL & PLRM & PQRM\\\hline
Surface area & 12.479 & 17.240 & \cellcolor{gray!57}12.438\\\hline\hline
\multicolumn{4}{||c||}{Cyanobacteria biomass vs total phosphorus}\\\hline
Methods & BL & PLRM & PQRM\\\hline
Surface area & 3.456 & 3.956 & \cellcolor{gray!57}2.951\\
   \hline\hline
\end{tabular}
\end{table}

\section{Discussion and conclusion} \label{sect:discon}

We applied a suite of statistical methods to estimate
ecological thresholds in two case studies. We proposed
to use the nonparametric regression loess to provide
a subjective idea of the number and positions
of the breakpoints. Having obtained initial values of the breakpoints
from loess, the piecewise linear regression model was used to estimate
the breakpoints based on the conditional mean of the dependent variable
given the independent variable. Finally,
a suite of estimates of the breakpoints for a range of quantiles
summarizing the full distribution of response values
was obtained using the piecewise linear quantile regression
method. Whereas the piecewise linear
regression model provides standard errors and confidence
intervals for the breakpoints (indicating the precision of the coefficients),
the piecewise linear quantile regression model generated narrower confidence intervals
for the breakpoints. The confidence intervals indicate the breakpoints'
most likely lower and upper bounds. Furthermore, we used the three
methods to generate prediction bands for response variable indicating
an index of biological condition given a predictor variable representing
environmental disturbance.

\subsection{Fish index of biotic integrity versus agricultural stress}

Bootstrapped loess (BL), piecewise linear
regression (PLRM) and the piecewise linear quantile
regression model (PQRM) all identified $2$
breakpoints (agricultural stress thresholds) in the
fish IBI vs agricultural stress dataset, all in
the same locations along the stress gradient. Of
the two methods from which confidence intervals
could be identified, the $80\%$ CI of the PQRM were
$59.30\%$ and $93.55\%$ as wide as those of the PLRM
breakpoints, for the lower and upper thresholds,
respectively.  Similarly, the PQRM estimates
produced the narrowest prediction bands,
enveloping an area only $72.15\%$ of that produced
by the PLRM. The BL envelope was similar to the
PQRM, bounding an area that was $72.38\%$ of the PLRM
area. 

The piecewise linear regression approach identified two breakpoints
along the agricultural stress gradient, at values $0.263$ and $0.488$.
The breakpoints were significantly different from each other and represented marked 
discontinuities in the index of biotic integrity-agricultural stress relationship.
Fish IBI scores were independent of agricultural stress
over the stress range from $0$ to $0.263$, but were a significant negative function of increasing
agricultural stress from $0.263$ to $0.488$. Fish IBI was also independent of agricultural stress
at stress values greater than $0.488$.
The $80\%$ confidence interval using the piecewise linear quantile regression model
for the first and second breakpoints are $(0.233, 0.284)$ and $(0.448, 564)$, respectively.
That is we are $80\%$ confident that the first breakpoint is situated between $0.233$ and $0.284$
and that the second breakpoint is located between $0.448$ and $0.564$.

Tests for departure of the $3$ segments of the fish
IBI vs agricultural stress regression line
indicated that fish IBI score was a negative
function of agricultural stress only between the
two stress thresholds. Below the lower threshold,
fish IBI scores was variable, but consistently
greater than 45 (the intercept of the first segment
of the $10th$ quantile of the PQRM). Above the upper
threshold, fish IBI scores were similarly variable
but consistently less than 38 (the maximum
interpolated value of the 90th quantile; Figure \ref{fig:QRM_IBI}).

The ecological and environmental management
implications of this interpretation are
significant. The models suggest that agricultural
stress values less than $0.196$ (lower bound of the
$95\%$ CI for the lower breakpoint) has no
detectable influence on fish community condition,
relative to the range of natural variation. In
contrast, under high levels of agricultural stress
($>0.585$; upper bound of the $95\%$ CI for the
higher breakpoint) management practices that
slightly reduce agricultural effects are unlikely
to improve fish community condition as expressed in
IBI scores. Agricultural changes to watersheds
draining into coastal wetlands are only likely to
influence fish community condition in wetlands with
stress scores between the two breakpoints.

\subsection{Cyanobacterial biomass versus total phosphorus}

The relative effectiveness of the $3$ regression
models (BL, PLRM and PQRM) for the cyanobacterial
biomass versus total phosphorus relationship matched
those observed for the fish IBI versus agricultural
stress analysis. Once again, the PQRM-derived
$80\%$ confidence intervals for the two thresholds were
narrower than the intervals derived from the PLRM
(only $67.49\%$ and $28.63\%$ of the PLRM-derived widths
for the lower and upper TP thresholds,
respectively).  As was the case for the fish IBI
dataset, the PLRM model produced the widest $80\%$
prediction envelop over the range of total
phosphorus concentrations. In this case, however,
the area of the $80\%$ prediction band estimated by
the PQRM was markedly narrower ($74.60\%$ of the PLRM
area) than the band estimated by the BL ($87.36\%$ of
the PLRM area). 

Our tests for the presence of ecological breakpoints in the data
relating cyanobacterial biomass to total phosphorus also identified two
statistically significant breakpoints, corresponding to
log of total phosphorus of $1.212$ ($16.293$ $\mu$g/L) and
$1.624$ ($42.073$ $\mu$g/L), respectively.
Cyanobacterial biomass increases slowly below log of total phosphorus
concentration of $1.212$, sharply between $1.212$ to $1.624$,
and slowly at log of total phosphorus concentration greater than $1.624$.
The $95\%$ confidence intervals using the piecewise linear 
quantile regression model for the first and second
breakpoints are $(1.026, 1.399)$ and $(1.418, 1.830)$, respectively.
That is, we are $95\%$ confident that the first breakpoint occurs
between $1.026$ and $1.399$, and the second breakpoint occurs
between $1.418$ and $1.830$.

The slopes of the $3$ segments of the
cyanobacterial biomass versus total phosphorus
relationship were all significantly
greater than zero, indicating that CB is associated
with the amount of TP across the entire nutrient
gradient. 
However, both the range of variation in CB and the slope of the relationship is much steeper over the range of TP concentrations between about $12$ ($10^{1.066}$) and $46$ ($10^{1.662}$) $\mu$g/L.
Below the lower threshold, CB is consistently less than about $100$  ($10^2$) $\mu$g/L, whereas above the upper threshold, CB is predicted to be greater than about $320$ ($10^{2.5}$) $\mu$g/L, ranging to as least $3\times 10^3$ at the upper limit of observed TP concentrations (Figure
\ref{fig:QRM_Cyan}).
The piecewise quantile regression analysis
lends support to both interpretations of the CB
versus TP relationship.  Biomass was a monotonically
increasing function of total phosphorus as
interpreted by \cite{Dillon:1974} and
\cite{Beaulieu:2014}. 
But the identification of
$3$ significantly different segments of the
relationship separated by breakpoints in the
nutrient gradient supports the sigmoidal
interpretation of the relationship \citep{Watson:1992,Filstrup:2014}. 
Clearly, the
identification of breakpoints indicates that the
relationship is more complicated than that implied
by linear regression of the log-transformed data.

The $80\%$ and $95\%$ bootstrap prediction bands using loess
for the fish index of biotic integrity and agricultural stress data (Figure \ref{Loess_CB_Bootstrap}) exhibit more uncertainty
than that of the cyanobacteria biomass and total phosphorus data (Figure \ref{Loess_PB_Bootstrap_Cyan})).
The prediction bands for the
fish index of biotic integrity and agricultural stress data
are more uncertain as this dataset contains $30$ data
points, which is much smaller than the $150$ data points
for the cyanobacteria biomass dataset.
This shows that the methodology is sensitive
to the sizes of the data sets.
Overall, the prediction
bands using the proposed bootstrap
method provides a good idea of the
spread of the future data points. 

The wide prediction bands generated using the piecewise linear regression model
derived for the fish index of biotic integrity and agricultural stress scores
reflect the limited number of observations ($n = 30$; Figure \ref{PLRM_Pred_Int})
and the number of degreess of freedom needed to fit the $6$ parameters of the model.
In contrast,
the fitting of multiple parameters was less inconsequential
to the width of prediction bands for the cyanobacteria
biomass
and total phosphorus concentration ($n = 150$; Figure \ref{PLRM_Pred_Int_Cyan}).

Sample size limitation exerted similar effects on the precision of prediction bands (Figure \ref{fig:QRM_IBI} \& \ref{fig:QRM_Cyan})
derived from the piecewise linear quantile regression.
In these cases, the effective sample size is a function of the quantile analyzed. As a result, the $80\%$ percentile
prediction bands are more erratic than the prediction bands generated by the 
bootstrapped loess and piecewise linear regression
models. However, as is true for all statistical methods, greater amounts of data would improve the precision of prediction bands.

The upper and lower bounds of prediction intervals (Figures \ref{Loess_CB_Bootstrap}, \ref{PLRM_Pred_Int}, \ref{Loess_PB_Bootstrap_Cyan} and \ref{PLRM_Pred_Int_Cyan})
derived from the bootstrapped loess and piecewise linear regression models reflect the relationships that are evident
in the central tendency of the data. These bounds do not reflect the relationships
that are present in the conditional quantiles of the data.
One advantage
of the prediction
bands obtained from the piecewise linear quantile regression is that the upper
and lower bounds (Figures \ref{fig:QRM_IBI} and \ref{fig:QRM_Cyan}) reflect the relationships in the
conditional quantiles of the data.

We recommend using the piecewise linear quantile regression model
to estimate the breakpoints as it provides the narrowest confidence intervals of the breakpoints.
Furthermore, the
piecewise linear quantile regression model provides
the smallest width of the prediction band for the 
data points.

\section*{Acknowledgments}

We thank Karen Fung
for her guidance and helpful comments on an early draft
of this manuscript. 
We also thank Sue Watson for informative discussions
on Chlorophyll-phosphorus relationships and Yakuta Bhagat, Donald J Uzarski and Lucinda Johnson for discussions about the shape of environmental stress biological response relationships. 
This research was supported by a grant from the 
Natural Sciences and Engineering Research Council of Canada to JJHC.


\section*{References}

\bibliography{mybibfile_v2}

\begin{thebibliography}{62}
\providecommand{\natexlab}[1]{#1}
\providecommand{\url}[1]{\texttt{#1}}
\expandafter\ifx\csname urlstyle\endcsname\relax
  \providecommand{\doi}[1]{doi: #1}\else
  \providecommand{\doi}{doi: \begingroup \urlstyle{rm}\Url}\fi

\bibitem[Austin(2007)]{Austin:2007}
M.~Austin.
\newblock Species distribution models and ecological theory: A critical
  assessment and some possible new approaches.
\newblock \emph{Ecological Modelling}, 200\penalty0 (1):\penalty0 1 -- 19,
  2007.

\bibitem[Beaulieu et~al.(2014)Beaulieu, Pick, Palmer, Watson, Winter, Zurawell,
  and Gregory-Eaves]{Beaulieu:2014}
M.~Beaulieu, F.~Pick, M.~Palmer, S.~Watson, J.~Winter, R.~Zurawell, and
  I.~Gregory-Eaves.
\newblock Comparing predictive cyanobacterial models from temperate regions.
\newblock \emph{Canadian Journal of Fisheries and Aquatic Sciences},
  71\penalty0 (12):\penalty0 1830--1839, 2014.

\bibitem[Bhagat et~al.(2007)Bhagat, Ciborowski, Johnson, Uzarski, Burton,
  Timmermans, and Cooper]{Bhagat:2007}
Y.~Bhagat, J.J.H. Ciborowski, L.B. Johnson, D.G. Uzarski, T.M. Burton, S.T.A.
  Timmermans, and M.J. Cooper.
\newblock Testing a fish index of biotic integrity for responses to different
  stressors in great lakes coastal wetlands.
\newblock \emph{Journal of Great Lakes Research}, 33:\penalty0 224 -- 235,
  2007.

\bibitem[Bissinger et~al.(2008)Bissinger, Montagnes, Sharples, and
  Atkinson]{Bissinger:2008}
J.E. Bissinger, D.J.S. Montagnes, J.~Sharples, and D.~Atkinson.
\newblock Predicting marine phytoplankton maximum growth rates from
  temperature: Improving on the eppley curve using quantile regression.
\newblock \emph{Limnology and Oceanography}, 53\penalty0 (2):\penalty0
  487--493, 2008.

\bibitem[Brazner and Beals(1997)]{Brazner:1997}
J.C. Brazner and E.W. Beals.
\newblock Patterns in fish assemblages from coastal wetland and beach habitats
  in green bay, lake michigan: a multivariate analysis of abiotic and biotic
  forcing factors.
\newblock \emph{Canadian Journal of Fisheries and Aquatic Sciences},
  54\penalty0 (8):\penalty0 1743--1761, 1997.

\bibitem[Brenden et~al.(2008)Brenden, Wang, and Su]{Brenden:2008}
T.O. Brenden, L.~Wang, and Z.~Su.
\newblock Quantitative identification of disturbance thresholds in support of
  aquatic resource management.
\newblock \emph{Environmental Management}, 42:\penalty0 821–832, 2008.

\bibitem[Bullerjahn et~al.(2016)Bullerjahn, McKay, Davis, Baker, Boyer,
  D’Anglada, Doucette, Ho, Irwin, Kling, Kudela, Kurmayer, Michalak, Ortiz,
  Otten, Paerl, Qin, Sohngen, Stumpf, Visser, and Wilhelm]{Bullerjahn:2016}
G.S. Bullerjahn, R.M. McKay, T.W. Davis, D.B. Baker, G.L. Boyer, L.V.
  D’Anglada, G.J. Doucette, J.C. Ho, E.G. Irwin, C.L. Kling, R.M. Kudela,
  R.~Kurmayer, A.M. Michalak, J.D. Ortiz, T.G. Otten, H.W. Paerl, B.~Qin, B.L.
  Sohngen, R.P. Stumpf, P.M. Visser, and S.W. Wilhelm.
\newblock Global solutions to regional problems: Collecting global expertise to
  address the problem of harmful cyanobacterial blooms. a lake erie case study.
\newblock \emph{Harmful Algae}, 54:\penalty0 223 -- 238, 2016.

\bibitem[Bunea et~al.(1999)Bunea, Guttorp, and Richardson]{Bunea:1999}
F.~Bunea, P.~Guttorp, and T.~Richardson.
\newblock Ecological indices and graphical modeling of factors influencing
  benthic populations in streams.
\newblock Technical report, NRCSE Technical Report Series, 1999.

\bibitem[Cade and Noon(2003)]{Cade:2003}
B.S. Cade and B.R. Noon.
\newblock A gentle introduction to quantile regression for ecologists.
\newblock \emph{Frontiers in Ecology and the Environment}, 1\penalty0
  (8):\penalty0 412--420, 2003.

\bibitem[Cade et~al.(1999)Cade, Terrell, and Schroeder]{Cade:1999}
B.S. Cade, J.W. Terrell, and R.L. Schroeder.
\newblock Estimating effects of limiting factors with regression quantiles.
\newblock \emph{Ecology}, 80:\penalty0 311--323, 1999.

\bibitem[Cade et~al.(2005)Cade, Noon, and Flather]{Cade:2005}
B.S. Cade, B.R. Noon, and C.H. Flather.
\newblock Quantile regression reveals hidden bias and uncertainty in habitat
  models.
\newblock \emph{Ecology}, 86\penalty0 (3):\penalty0 786--800, 2005.

\bibitem[Campos and Vasconcelos(2010)]{Campos:2010}
A.~Campos and V.~Vasconcelos.
\newblock Molecular mechanisms of microcystin toxicity in animal cells.
\newblock \emph{International Journal of Molecular Sciences}, 11\penalty0
  (1):\penalty0 268--287, 2010.

\bibitem[Chow-Fraser et~al.(1994)Chow-Fraser, Trew, Findlay, and
  Stainton]{Chow-Fraser:1994}
P.~Chow-Fraser, D.O. Trew, D.~Findlay, and M.~Stainton.
\newblock A test of hypotheses to explain the sigmoidal relationship between
  total phosphorus and chlorophyll a concentrations in canadian lakes.
\newblock \emph{Canadian Journal of Fisheries and Aquatic Sciences},
  51\penalty0 (9):\penalty0 2052--2065, 1994.

\bibitem[Cleveland(1979)]{Cleveland:1979}
W.S. Cleveland.
\newblock Robust locally weighted regression and smoothing scatterplots.
\newblock \emph{Journal of the American Statistical Association}, 74\penalty0
  (368):\penalty0 829--836, 1979.

\bibitem[Crosbie and Chow-Fraser(1999)]{Crosbie:1999}
B.~Crosbie and P.~Chow-Fraser.
\newblock Percentage land use in the watershed determines the water and
  sediment quality of 22 marshes in the great lakes basin.
\newblock \emph{Canadian Journal of Fisheries and Aquatic Sciences},
  56\penalty0 (10):\penalty0 1781--1791, 1999.

\bibitem[Danz et~al.(2005)Danz, Regal, Niemi, Brady, Hollenhorst, Johnson,
  Host, Hanowski, Johnston, Brown, Kingston, and Kelly]{Danz:2005}
N.P. Danz, R.R. Regal, G.J. Niemi, V.J. Brady, T.~Hollenhorst, L.B. Johnson,
  G.E. Host, J.M. Hanowski, C.A. Johnston, T.~Brown, J.~Kingston, and J.R.
  Kelly.
\newblock Environmentally stratified sampling design for the development of
  great lakes environmental indicators.
\newblock \emph{Environmental Monitoring and Assessment}, 102\penalty0
  (1):\penalty0 41--65, 2005.

\bibitem[Davison and Hinkley(1997)]{Davison:1997}
A.C. Davison and D.V. Hinkley.
\newblock \emph{Bootstrap Methods and Their Application}.
\newblock Cambridge Series in Statistical and Probabilistic Mathematics.
  Cambridge University Press, 1997.

\bibitem[Dillon and Rigler(1974)]{Dillon:1974}
P.J. Dillon and F.H. Rigler.
\newblock The phosphorus-chlorophyll relationship in lakes1,2.
\newblock \emph{Limnology and Oceanography}, 19\penalty0 (5):\penalty0
  767--773, 1974.

\bibitem[Dillon and Rigler(1975)]{Dillon:1975}
P.J. Dillon and F.H. Rigler.
\newblock A simple method for predicting the capacity of a lake for development
  based on lake trophic status.
\newblock \emph{Journal of the Fisheries Research Board of Canada}, 32\penalty0
  (9):\penalty0 1519--1531, 1975.

\bibitem[Dodds et~al.(2010)Dodds, Clements, Gido, Hilderbrand, and
  King]{Dodds:2010}
W.K. Dodds, W.H. Clements, K.~Gido, R.H. Hilderbrand, and R.S. King.
\newblock Thresholds, breakpoints, and nonlinearity in freshwaters as related
  to management.
\newblock \emph{Journal of the North American Benthological Society},
  29\penalty0 (3):\penalty0 988--997, 2010.

\bibitem[Downing et~al.(2001)Downing, Watson, and McCauley]{Downing:2001}
J.A. Downing, S.B. Watson, and E.~McCauley.
\newblock Predicting cyanobacteria dominance in lakes.
\newblock \emph{Canadian Journal of Fisheries and Aquatic Sciences},
  58\penalty0 (10):\penalty0 1905--1908, 2001.

\bibitem[Efron and Tibshirani(1994)]{Efron:1994}
B.~Efron and R.J. Tibshirani.
\newblock \emph{An Introduction to the Bootstrap}.
\newblock Chapman \& Hall/CRC Monographs on Statistics \& Applied Probability.
  Taylor \& Francis, 1994.

\bibitem[Fahrig(2001)]{Fahrig:2001}
L.~Fahrig.
\newblock How much habitat is enough?
\newblock \emph{Biological Conservation}, 100\penalty0 (1):\penalty0 65 -- 74,
  2001.

\bibitem[Ficetola and Denoël(2009)]{Ficetola:2009}
G.F. Ficetola and M.~Denoël.
\newblock Ecological thresholds: an assessment of methods to identify abrupt
  changes in species–habitat relationships.
\newblock \emph{Ecography}, 32\penalty0 (6):\penalty0 1075--1084, 2009.

\bibitem[Filstrup et~al.(2014)Filstrup, Wagner, Soranno, Stanley, Stow,
  Webster, and Downing]{Filstrup:2014}
C.T. Filstrup, T.~Wagner, P.A. Soranno, E.H. Stanley, C.A. Stow, K.E. Webster,
  and J.A. Downing.
\newblock Regional variability among nonlinear chlorophyll—phosphorus
  relationships in lakes.
\newblock \emph{Limnology and Oceanography}, 59\penalty0 (5):\penalty0
  1691--1703, 2014.

\bibitem[Francesco~Ficetola and Denoël(2009)]{Francesco:2009}
G.~Francesco~Ficetola and M.~Denoël.
\newblock Ecological thresholds: an assessment of methods to identify abrupt
  changes in species–habitat relationships.
\newblock \emph{Ecography}, 32\penalty0 (6):\penalty0 1075--1084, 2009.

\bibitem[Frey(1977)]{Frey:1977}
D.G. Frey.
\newblock Biological integrity of water - an historic approach.
\newblock In R.K. Ballantine and L.J. Guarraia, editors, \emph{The Integrity of
  Water}, pages 127--140. U.S. Environmental Protection Agency, 1977.

\bibitem[Granato(2007)]{Granato:2006}
G.E. Granato.
\newblock \emph{Kendall-Theil Robust Line ( KTRLine – version 1 . 0 ) — A
  Visual Basic Program for Calculating and Graphing Robust Nonparametric
  Estimates of Linear-Regression Coefficients Between Two Continuous Variables
  Techniques}.
\newblock Reston, Va: U.S. Dept. of the Interior, U.S. Geological Survey, 2007.
\newblock URL \url{http://purl.access.gpo.gov/GPO/LPS97008}.

\bibitem[H{\"a}rdle and Bowman(1988)]{Hardle:1988}
W.~H{\"a}rdle and A.W. Bowman.
\newblock Bootstrapping in nonparametric regression: Local adaptive smoothing
  and confidence bands.
\newblock \emph{Journal of the American Statistical Association}, 83\penalty0
  (401):\penalty0 102--110, 1988.

\bibitem[Horning(2012)]{Horning:2012}
M.~Horning.
\newblock Constraint lines and performance envelopes in behavioral physiology:
  the case of the aerobic dive limit.
\newblock \emph{Frontiers in Physiology}, 3:\penalty0 381, 2012.
\newblock \doi{10.3389/fphys.2012.00381}.

\bibitem[Host et~al.(2005)Host, Schuldt, Ciborowski, Johnson, Hollenhorst, and
  Richards]{Host:2005}
G.E. Host, J.~Schuldt, J.J.H. Ciborowski, L.B. Johnson, T.~Hollenhorst, and
  C.~Richards.
\newblock Use of gis and remotely sensed data for a priori identification of
  reference areas for great lakes coastal ecosystems.
\newblock \emph{International Journal of Remote Sensing}, 26\penalty0
  (23):\penalty0 5325--5342, 2005.

\bibitem[Johnson(2013)]{Johnson:2013}
C.J. Johnson.
\newblock Identifying ecological thresholds for regulating human activity:
  Effective conservation or wishful thinking?
\newblock \emph{Biological Conservation}, 168:\penalty0 57 -- 65, 2013.

\bibitem[Karanth et~al.(2004)Karanth, Nichols, Kumar, Link, and
  Hines]{Karanth:2004}
K.U. Karanth, J.D. Nichols, N.S. Kumar, W.A. Link, and J.E. Hines.
\newblock Tigers and their prey: predicting carnivore densities from prey
  abundance.
\newblock \emph{Proceedings of the National Academy of Sciences of the United
  States of America}, 101\penalty0 (14):\penalty0 4854--4858, 2004.

\bibitem[Karr and Chu(1998)]{Karr:1998}
J.R. Karr and E.W. Chu.
\newblock \emph{Restoring Life in Running Waters: Better Biological
  Monitoring}.
\newblock Island Press, 1998.

\bibitem[Kilgour et~al.(1998)Kilgour, Somers, and Matthews]{KILGOUR:1998}
B.W. Kilgour, K.M. Somers, and D.E. Matthews.
\newblock Using the normal range as a criterion for ecological significance in
  environmental monitoring and assessment.
\newblock \emph{Écoscience}, 5\penalty0 (4):\penalty0 542--550, 1998.

\bibitem[King and Baker(2011)]{King:2011}
R.S. King and M.E. Baker.
\newblock An alternative view of ecological community thresholds and
  appropriate analyses for their detection: comment.
\newblock \emph{Ecological Applications}, 21\penalty0 (7):\penalty0 2833--2839,
  2011.

\bibitem[Koenker(2005)]{Koenker:2005}
R.~Koenker.
\newblock \emph{Quantile Regression}.
\newblock Cambridge: Cambridge University Press, 2005.

\bibitem[Koenker et~al.(2018)Koenker, Portnoy, Ng, Zeileis, Grosjean, and
  Ripley]{Quantreg:2018}
R.~Koenker, S.~Portnoy, P.T. Ng, A.~Zeileis, P.~Grosjean, and B.D. Ripley.
\newblock \emph{Quantile Regression}, 2018.
\newblock R package version 5.36.

\bibitem[Labine and Minuk(2009)]{Labine:2009}
M.A. Labine and G.Y. Minuk.
\newblock Cyanobacterial toxins and liver diseasethis article is one of a
  selection of papers published in a special issue celebrating the 125th
  anniversary of the faculty of medicine at the university of manitoba.
\newblock \emph{Canadian Journal of Physiology and Pharmacology}, 87\penalty0
  (10):\penalty0 773--788, 2009.

\bibitem[Lone et~al.(2015)Lone, Koiri, and Bhide]{Lone:2015}
Y.~Lone, K.R. Koiri, and M.~Bhide.
\newblock An overview of the toxic effect of potential human carcinogen
  microcystin-lr on testis.
\newblock \emph{Toxicology Reports}, 2:\penalty0 289 -- 296, 2015.

\bibitem[McClain and Rex(2001)]{McClain:2001}
C.~McClain and M.~Rex.
\newblock The relationship between dissolved oxygen concentration and maximum
  size in deep-sea turrid gastropods: an application of quantile regression.
\newblock \emph{Marine Biology}, 139\penalty0 (4):\penalty0 681--685, 2001.

\bibitem[Mosteller and Tukey(1977)]{Mosteller:1977}
F.~Mosteller and J.W. Tukey.
\newblock \emph{Data Analysis and Regression: A Second Course in Statistics}.
\newblock Addison-Wesley series in behavioral science. Addison-Wesley
  Publishing Company, 1977.

\bibitem[Muggeo(2015)]{Segmented:2015}
V.M.R. Muggeo.
\newblock \emph{segmented: Regression Models with Breakpoints/Changepoints
  Estimation}, 2015.
\newblock R package version 0.5-1.4.

\bibitem[Pick(2016)]{Pick:2016}
F.R. Pick.
\newblock Blooming algae: a canadian perspective on the rise of toxic
  cyanobacteria.
\newblock \emph{Canadian Journal of Fisheries and Aquatic Sciences},
  73\penalty0 (7):\penalty0 1149--1158, 2016.

\bibitem[Planque and Buffaz(2008)]{Planque:2008}
B.~Planque and L.~Buffaz.
\newblock Quantile regression models for fish recruitment–environment
  relationships: four case studies.
\newblock \emph{Marine Ecology Progress Secries}, 357:\penalty0 213--223, 2008.

\bibitem[Qian and Miltner(2015)]{Qian:2015}
S.S. Qian and R.J. Miltner.
\newblock A continuous variable bayesian networks model for water quality
  modeling.
\newblock \emph{Environ. Model. Softw.}, 69\penalty0 (C):\penalty0 14--22, July
  2015.
\newblock ISSN 1364-8152.

\bibitem[Qian et~al.(2003)Qian, King, and Richardson]{Qian:2003}
S.S. Qian, R.S. King, and C.J. Richardson.
\newblock Two statistical methods for the detection of environmental
  thresholds.
\newblock \emph{Ecological Modelling}, 166\penalty0 (1):\penalty0 87 -- 97,
  2003.

\bibitem[{R Core Team}(2017)]{Rcite}
{R Core Team}.
\newblock \emph{R: A Language and Environment for Statistical Computing}.
\newblock R Foundation for Statistical Computing, Vienna, Austria, 2017.
\newblock URL \url{https://www.R-project.org}.

\bibitem[Reynolds and Walsby(1975)]{Reynolds:1975}
C.S. Reynolds and A.E. Walsby.
\newblock Water-blooms.
\newblock \emph{Biological Reviews}, 50\penalty0 (4):\penalty0 437--481, 1975.

\bibitem[Roegner et~al.(2014)Roegner, Brena, González-Sapienza, and
  Puschner]{Roegner:2014}
A.F. Roegner, B.~Brena, G.~González-Sapienza, and B.~Puschner.
\newblock Microcystins in potable surface waters: toxic effects and removal
  strategies.
\newblock \emph{Journal of Applied Toxicology}, 34\penalty0 (5):\penalty0
  441--457, 2014.

\bibitem[Scharf et~al.(1998)Scharf, Juanes, and Sutherland]{Scharf:1998}
F.S. Scharf, F.~Juanes, and M.~Sutherland.
\newblock Inferring ecological relationships from edges of scatter diagrams:
  Comparison of regression techniques.
\newblock \emph{Ecology}, 79\penalty0 (2):\penalty0 448--460, 1998.

\bibitem[Seber and Wild(2003)]{Seber:2003}
G.A.F. Seber and C.J. Wild.
\newblock \emph{Nonlinear Regression}.
\newblock Wiley Series in Probability and Statistics. John Wiley \& Sons, Inc.,
  Hoboken, New Jersey, 2003.

\bibitem[Shea and Vecchione(2002)]{Shea:2002}
E.~Shea and M.~Vecchione.
\newblock Quantification of ontogenetic discontinuities in three species of
  oegopsid squids using model ii piecewise linear regression.
\newblock \emph{Marine Biology}, 140\penalty0 (5):\penalty0 971--979, 2002.

\bibitem[Stoddard et~al.(2006)Stoddard, Larsen, Hawkins, Johnson, and
  Norris]{Stoddard:2006}
J.L. Stoddard, D.P. Larsen, C.P. Hawkins, R.K. Johnson, and R.H. Norris.
\newblock Setting expectations for the ecological condition of streams: The
  concept of reference condition.
\newblock \emph{Ecological Applications}, 16\penalty0 (4):\penalty0 1267--1276,
  2006.

\bibitem[Stow and Cha(2013)]{Stow:2013}
C.A. Stow and Y.~Cha.
\newblock Are chlorophyll a–total phosphorus correlations useful for
  inference and prediction?
\newblock \emph{Environmental Science \& Technology}, 47\penalty0 (8):\penalty0
  3768--3773, 2013.

\bibitem[Svendsen et~al.(2018)Svendsen, Andersen, Hansen, and
  Steffensen]{Sondergaard:2018}
M.B.S. Svendsen, N.R. Andersen, P.J. Hansen, and J.F. Steffensen.
\newblock Effects of harmful algal blooms on fish: Insights from prymnesium
  parvum.
\newblock \emph{Fishes}, 3\penalty0 (1), 2018.
\newblock \doi{10.3390/fishes3010011}.

\bibitem[Terrell et~al.(1996)Terrell, Cade, Carpenter, and
  Thompson]{Terrell:1996}
J.W. Terrell, B.S. Cade, J.~Carpenter, and J.M. Thompson.
\newblock Modeling stream fish habitat limitations from wedge-shaped patterns
  of variation in standing stock.
\newblock \emph{Transactions of the American Fisheries Society}, 125\penalty0
  (1):\penalty0 104--117, 1996.

\bibitem[Toms and Lesperance(2003)]{Toms:2003}
J.D. Toms and M.L. Lesperance.
\newblock Piecewise regression: A tool for identifying ecological thresholds.
\newblock \emph{Ecology}, 84\penalty0 (8):\penalty0 2034--2041, 2003.

\bibitem[Trexler and Travis(1993)]{Trexler:1993}
J.C. Trexler and J.~Travis.
\newblock Nontraditional regression analyses.
\newblock \emph{Ecology}, 74\penalty0 (6):\penalty0 1629--1637, 1993.

\bibitem[Uzarski et~al.(2005)Uzarski, Burton, Cooper, Ingram, and
  Timmermans]{Uzarski:2005}
D.G. Uzarski, T.M. Burton, M.J. Cooper, J.W. Ingram, and S.T.A. Timmermans.
\newblock Fish habitat use within and across wetland classes in coastal
  wetlands of the five great lakes: Development of a fish-based index of biotic
  integrity.
\newblock \emph{Journal of Great Lakes Research}, 31:\penalty0 171 -- 187,
  2005.

\bibitem[Watson et~al.(1992)Watson, McCauley, and Downing]{Watson:1992}
S.~Watson, E.~McCauley, and J.A. Downing.
\newblock Sigmoid relationships between phosphorus, algal biomass, and algal
  community structure.
\newblock \emph{Canadian Journal of Fisheries and Aquatic Sciences},
  49\penalty0 (12):\penalty0 2605--2610, 1992.

\bibitem[Watson et~al.(1997)Watson, McCauley, and Downing]{Watson:1997}
S.B. Watson, E.~McCauley, and J.A. Downing.
\newblock Patterns in phytoplankton taxonomic composition across temperate
  lakes of differing nutrient status.
\newblock \emph{Limnology and Oceanography}, 42\penalty0 (3):\penalty0
  487--495, 1997.

\end{thebibliography}

\end{document}